\documentclass[fleqn,usenatbib]{mnras}

\usepackage[T1]{fontenc}

\usepackage{graphicx}	
\usepackage{amsmath}	
\usepackage{amssymb}	
\usepackage{multirow}
\usepackage{booktabs}
\usepackage{caption,subcaption}

\title[An HST survey of 33 T8 to Y1 brown dwarfs]{An HST survey of 33 T8 to Y1 brown dwarfs: NIR photometry and multiplicity of the coldest isolated objects}

\author[C. Fontanive et al.]{
Clémence Fontanive,$^{1}$\thanks{E-mail: \href{mailto:clemence.fontanive@umontreal.ca} {clemence.fontanive@umontreal.ca}}
Luigi R. Bedin,$^{2}$
Matthew De Furio,$^{3}$
Beth Biller,$^{4}$
Jay Anderson,$^{5}$
\newauthor Mariangela Bonavita,$^{6}$
Katelyn Allers,$^{7}$
Blake Pantoja$^{7}$
\\
$^{1}$Trottier Institute for Research on Exoplanets, Université de Montréal, Montréal, H3C 3J7, Québec, Canada\\
$^{2}$NAF-Osservatorio Astronomico di Padova, Vicolo dell’Osservatorio 5, I-35122 Padova, Italy\\
$^{3}$Department of Astronomy, University of Michigan, Ann Arbor, MI 48109, USA\\
$^{4}$SUPA, Institute for Astronomy, University of Edinburgh, Blackford Hill, Edinburgh EH9 3HJ, UK\\
$^{5}$Space Telescope Science Institute, 3700 San Martin Drive, Baltimore, MD 21218, USA\\
$^{6}$School of Physical Sciences, The Open University, Walton Hall, Milton Keynes MK7 6AA, UK\\
$^{7}$Department of Physics and Astronomy, Bucknell University, Lewisburg, PA 17837, USA
}

\date{Accepted 2023 September 17. Received 2023 May 15}

\pubyear{2023}

\begin{document}
\label{firstpage}
\pagerange{\pageref{firstpage}--\pageref{lastpage}}
\maketitle

\begin{abstract}
We present results from a Hubble Space Telescope imaging search for low-mass binary and planetary companions to 33 nearby brown dwarfs with spectral types of T8--Y1. Our survey provides new photometric information for these faint systems, from which we obtained model-derived luminosities, masses and temperatures. Despite achieving a deep sensitivity to faint companions beyond 0.2--0.5\arcsec, down to mass ratios of 0.4--0.7 outside $\sim$5~au, we find no companions to our substellar primaries. From our derived survey completeness, we place an upper limit of $f < 4.9\%$ at the 1-$\sigma$ level ($< 13.0$\% at the 2-$\sigma$ level) on the binary frequency of these objects over the separation range 1--1000~au and for mass ratios above $q=0.4$. Our results confirm that companions are extremely rare around the lowest-mass and coldest isolated brown dwarfs, continuing the marginal trend of decreasing binary fraction with primary mass observed throughout the stellar and substellar regimes. These findings support the idea that if a significant population of binaries exist around such low-mass objects, it should lie primarily below 2--3~au separations, with a true peak possibly located at even tighter orbital separations for Y dwarfs.
\end{abstract}




\section{Introduction}
\label{intro}

Multiplicity, as a direct outcome of formation, is a fundamental parameter in the study of populations of stars and brown dwarfs. Multiplicity studies can help disentangle different formation mechanisms that produce distinct binary outcomes. The binary properties (binary rate, orbital separation and mass ratio distributions) of field objects show a continuous decrease in binary frequency from $\sim$50\% for Sun-like stars \citep{Raghavan2010}, to $\sim$30\% for M dwarfs \citep{Ward-Duong2015,Winters2019} and $\sim$10--20\% for L-T brown dwarfs \citep{Close2003,Burgasser2006,Reid2006,Aberasturi2014}, with later-type binaries found to be more compact and with a clear tendency towards equal-mass systems. This observed smooth continuity in multiplicity properties, from massive stars, across the substellar limit, down to late-type brown dwarfs, points to a common formation mechanism between the stellar and substellar regimes \citep{Allen2007}.

\citet{Fontanive2018} placed the first statistically-robust constraints to date on the binary fraction of $\geq$T8 ultracool dwarfs ($T_\mathrm{eff}<800$~K), finding an inherently low binary rate ($8\pm6\%$), with a very tight binary separation distribution peaking around $\sim$3~au, and a $<$1\% binary rate beyond 10~au. While these results confirm previous trends, they remain severely limited by small sample sizes and low numbers of detections. In particular, the sample probed by \citet{Fontanive2018} only included 3 Y-type primaries ($T_\mathrm{eff}<500$~K), all Y0 brown dwarfs. Larger samples are thus still required to confirm whether these tendencies also hold for the very latest spectral types and lowest primary masses, where only high mass ratio binaries ($q\gtrsim0.7$) have been probed within 5--10~au separations \citep{Opitz2016,Fontanive2018}. 

Probing distinct stages in the lives of brown dwarfs is similarly critical to discriminate between primordial formation and subsequent dynamical evolution. A number of wide-separation, low-mass brown dwarf binaries have been discovered in young star-forming regions, including 2M1207 (25 and 5~M$_\mathrm{Jup}$, 40~au; \citealp{Chauvin2005}) and 2MASS~0441$+$2301 (20 and 10~M$_\mathrm{Jup}$, 15~au; \citealp{Todorov2010}) (see also recent discoveries by \citealp{DeFurio2022}). Yet more surprising systems like the 12-M$_\mathrm{Jup}$ tertiary component orbiting at 2000~au from the more massive 2MASS~0249$-$0557~AB binary (48 and 44~M$_\mathrm{Jup}$; \citealp{Dupuy2018}), or the recently-discovered Oph~98 system (15 and 8~M$_\mathrm{Jup}$, 200~au \citealp{Fontanive2021}) challenge these theories even further. 
The strong lack of binaries with such large orbital separations and uneven component masses within the evolved field population, compared to objects of similar masses in young associations, is seen as evidence that such weakly-bound system do not survive to field ages \citep{Burgasser2007,Biller2011}, providing valuable insights into the result of formation and evolutionary patterns.

Nonetheless, some rare examples of wide, late-type binaries with low mass ratios are  known to exist. For example, WISE~1217$+$1626 (T9$+$Y0, 12 and 8~M$_\mathrm{Jup}$) and WISE~J1711$+$3500 (T8$+$T9.5, 20 and 9~M$_\mathrm{Jup}$) have separations of $\sim$8--15~au \citep{Liu2012}, making them close analogues to 2M1207 and 2MASS~0441+2301 in terms of binary configurations. These systems are difficult to reconcile with most formation scenarios for brown dwarfs, that only allow tight binaries to survive (e.g., ejection scenario, \citealp{ReipurthClarke2001}; disc fragmentation and binary disruption, \citealp{GoodwinWhitworth2007}). Given that very few such low-mass primaries have been probed for binarity on these wide separations and down to such low mass ratios, it is unclear whether these peculiar discoveries are simply uncommon, or whether more field counterparts to the most extreme young systems exist. Nevertheless, no field counterparts to the most extreme young systems, with separations of hundreds of au, have been identified to date for brown dwarf primaries with masses below $\sim$40~M$_\mathrm{Jup}$. The discovery of such extreme systems at advanced ages would represent key elements to help us identify and further understand the mechanisms at play in the formation of the observed brown dwarf population. 

In this paper, we present \textit{Hubble Space Telescope} (\textit{HST}) observations of 33 T8--Y1 brown dwarfs to search for low-mass companions to these systems. With 9 Y dwarf primaries, this survey contains the largest Y-dwarf sample probed for multiplicity to date. The goals of this campaign are (1) to refine the binary statistics of the very latest-type objects, currently poorly constrained, (2) to confirm whether widely-separated (tens of au) low mass ratio systems are indeed more common around $>$T8 dwarfs than around their more massive, earlier-type counterparts, and (3) to search for the first extremely wide (hundreds of au) binary companion to a very late-type field-age primary. Collectively answering these questions will place strong constraints on the frequency of wide, planetary-mass companions at old ages, providing key insights into the formation and dynamical evolution of these systems.
We describe the sample and observations in Section~\ref{observations}. Sections~\ref{data_analysis:primaries} and \ref{data_analysis:binarity} present the data analyses of the science targets and the companion search, respectively. Statistical results of the multiplicity survey are presented in Section~\ref{stats}, and discussed in Section~\ref{discussion}. Our results are summarised in Section~\ref{conclusions}.

\section{HST/WFC3 Observations}
\label{observations}

\subsection{Sample Selection}
\label{sample}

The studied sample consists of 33 T8 and later-type nearby ($<30$~pc) field brown dwarfs, previously identified via the \textit{Wide-Field Infrared Survey Explorer} (\textit{WISE}; \citealp{Wright2010}). The sample includes the 22 unobserved targets from \textit{HST} snapshot program SNAP 12873 (PI Biller), which was published in \citet{Fontanive2018} for the 12 successfully observed systems. To these were added 11 brown dwarfs from \citet{Schneider2015} that hadn't already been targeted in a multiplicity search with comparable detection limits to the predicted sensitivity of the designed \textit{HST} program (GO 15201, PI Fontanive).

\begin{table*}
    \addtolength{\tabcolsep}{2pt}
    \renewcommand{\arraystretch}{1.2}
    \centering
    \caption{Probed sample of late-T and Y brown dwarfs.}
    \begin{small}
    \begin{tabular}{l c c c c c c c c}
    \hline\hline
        Target Name & Short Name & RA & DEC & Disc. & SpT & SpT & $\varpi_\mathrm{abs}$ & Astrom. \\
         & & (J2000) & (J2000) & Ref. & (IR) & Ref. & (mas) & Ref. \\
    \hline
        WISE J000517.48$+$373720.5  & WISE 0005$+$3737  & 00:05:17.48   & $+$37:37:20.5 & (1)   & T9.0    & (1)   & $126.9\pm2.1$ & (10) \\
        WISE J001505.87$-$461517.6  & WISE 0015$-$4615  & 00:15:05.88	& $-$46:15:17.7	& (2)   & T8.0    & (2)   & $75.2\pm2.4$  & (10) \\
        WISE J003231.09$-$494651.4  & WISE 0032$-$4946  & 00:32:31.09	& $-$49:46:51.5	& (2)   & T8.5    & (2)   & $60.8\pm2.5$  & (10) \\
        WISE J003829.05$+$275852.1  & WISE 0038$+$2758  & 00:38:29.06	& $+$27:58:52.1	& (1)   & T9.0    & (1)   & $88.2\pm2.0$  & (10) \\
        WISE J004945.61$+$215120.0  & WISE 0049$+$2151  & 00:49:46.09	& $+$21:51:20.4	& (1)   & T8.5    & (1)   & $140.4\pm2.1$ & (10) \\
        CFBDS J013302$+$023128      & CFBDS 0133$+$0231  & 01:33:02.48	& $+$02:31:28.9	& (3)   & T8.5    & (3)   & $53.1\pm2.6$  & (10) \\
        WISE J032517.69$-$385454.1  & WISE 0325$-$3854  & 03:25:17.69	& $-$38:54:54.1	& (1)   & T9.0    & (1)   & $60.2\pm3.5$  & (10) \\
        WISE J032504.33$-$504400.3  & WISE 0325$-$5044  & 03:25:04.52	& $-$50:44:03.0	& (4)   & T8.0    & (4)   & $36.7\pm2.7$  & (11) \\
        WISE J035000.32$-$565830.2  & WISE 0350$-$5658  & 03:50:00.33 	& $-$56:58:30.2 & (2)   & Y1.0    & (2)   & $176.4\pm2.3$ & (10) \\
        WISE J035934.06$-$540154.6  & WISE 0359$-$5401  & 03:59:34.07	& $-$54:01:54.6	& (2)   & Y0.0    & (2)   & $73.6\pm2.0$  & (10) \\
        WISE J040443.48$-$642029.9  & WISE 0404$-$6420  & 04:04:43.50	& $-$64:20:30.0	& (4)   & T9.0    & (4)   & $44.8\pm2.2$  & (10) \\
        WISE J041022.71$+$150248.4  & WISE 0410$+$1502  & 04:10:22.71	& $+$15:02:48.5	& (5)   & Y0.0    & (5)   & $151.3\pm2.0$ & (10) \\
        WISE J041358.14$-$475039.3  & WISE 0413$-$4750  & 04:13:58.14	& $-$47:50:39.3	& (1)   & T9.0    & (1)   & $50.7\pm3.3$  & (10) \\
        WISE J074457.25$+$562821.0  & WISE 0744$+$5628  & 07:44:57.25	& $+$56:28:21.0	& (6)   & T8.0    & (6)   & $65.3\pm2.0$  & (10) \\
        WISE J081117.81$-$805141.3  & WISE 0811$-$8051  & 08:11:17.82	& $-$80:51:41.4	& (1)   & T9.5    & (1)   & $99.1\pm7.7$  & (11,12) \\
        WISE J081220.04$+$402106.2  & WISE 0812$+$4021  & 08:12:20.04	& $+$40:21:06.3	& (1)   & T8.0    & (1)   & $34.3\pm2.71$  & (10) \\
        WISE J085716.24$+$560407.6  & WISE 0857$+$5604  & 08:57:16.25	& $+$56:04:07.7	& (6)   & T8.0    & (6)   & $85.3\pm2.1$  & (10) \\
        WISE J094305.98$+$360723.5  & WISE 0943$+$3607  & 09:43:05.99	& $+$36:07:23.6	& (7)   & T9.5    & (7)   & $97.1\pm2.9$  & (10) \\
        WISE J105130.01$-$213859.7  & WISE 1051$-$2138  & 10:51:30.02	& $-$21:38:59.7	& (1)   & T8.5    & (8)   & $64.0\pm2.3$  & (10) \\
        WISE J120604.38$+$840110.6  & WISE 1206$+$8401  & 12:06:04.39	& $+$84:01:10.7	& (4)   & Y0.0    & (4)   & $84.7\pm2.1$  & (10) \\
        WISE J131833.98$-$175826.5  & WISE 1318$-$1758  & 13:18:33.98	& $-$17:58:26.5	& (1)   & T8.0    & (8)   & $63.5\pm2.2$  & (10) \\
        WISE J154151.65$-$225024.9  & WISE 1541$-$2250  & 15:41:51.66	& $-$22:50:25.0	& (5)   & Y1.0    & (4)   & $168.6\pm2.2$ & (13) \\
        WISE J163940.83$-$684738.6  & WISE 1639$-$6847  & 16:39:40.84	& $-$68:47:38.6	& (9)   & Y0.0p   & (4)   & $211.1\pm0.6$ & (14) \\
        WISE J173835.53$+$273259.0  & WISE 1738$+$2732  & 17:38:35.53	& $+$27:32:59.1	& (5)   & Y0.0    & (5)   & $130.9\pm2.1$ & (10) \\
        WISE J201920.76$-$114807.5  & WISE 2019$-$1148  & 20:19:20.77	& $-$11:48:07.6	& (1)   & T8.0    & (1)   & $79.9\pm2.7$  & (10) \\
        WISE J205628.91$+$145953.2  & WISE 2056$+$1459  & 20:56:28.92	& $+$14:59:53.2	& (5)   & Y0.0    & (5)   & $140.8\pm2.0$ & (10) \\
        WISE J210200.15$-$442919.5  & WISE 2102$-$4429  & 21:02:00.16	& $-$44:29:19.5	& (2)   & T9.0    & (2)   & $92.9\pm1.9$  & (11,12) \\
        WISE J221216.33$-$693121.6  & WISE 2212$-$6931  & 22:12:16.27	& $-$69:31:21.6	& (4)   & T9.0    & (4)   & $80.6\pm1.9$  & (10) \\
        WISE J225540.75$-$311842.0  & WISE 2255$-$3118  & 22:55:40.75	& $-$31:18:42.1	& (6)   & T8.0    & (6)   & $72.8\pm3.5$  & (10) \\
        WISE J231336.38$-$803700.1  & WISE 2313$-$8037  & 23:13:36.38	& $-$80:37:00.2	& (6)   & T8.0    & (6)   & $92.6\pm2.2$  & (10) \\
        WISE J232519.53$-$410535.0  & WISE 2325$-$4105  & 23:25:19.54	& $-$41:05:35.1	& (6)   & T9.0p   & (6)   & $108.4\pm3.7$ & (11,12) \\
        WISE J233226.49$-$432510.6  & WISE 2332$-$4325  & 23:32:26.50	& $-$43:25:10.6	& (2)   & T9.0    & (2)   & $61.1\pm2.1$  & (10) \\
        WISE J235402.77$+$024015.0  & WISE 2354$+$0240  & 23:54:02.79 	& $+$02:40:14.1	& (4)   & Y1.0    & (4)   & $130.6\pm3.3$ & (10) \\
    \hline \\ [-2.5ex]
    \multicolumn{9}{l}{
    \begin{minipage}{0.94\textwidth}
        \textbf{Notes.} Parallax measurements are all trigonometric.\\
        \textbf{References.}
        (1) \citet{Mace2013a};
        (2) \citet{Kirkpatrick2012};
        (3) \citet{Albert2011};
        (4) \citet{Schneider2015};
        (5) \citet{Cushing2011};
        (6) \citet{Kirkpatrick2011};
        (7) \citet{Cushing2014};
        (8) \citet{Martin2018};
        (9) \citet{Tinney2012};
        (10) \citet{Kirkpatrick2021};
        (11) \citet{Kirkpatrick2019};
        (12) \citet{Tinney2014};
        (13) \citet{BedinFontanive2018};
        (14) \citet{Fontanive2021}.
    \end{minipage}}
    \label{t:targets_sample}
    \end{tabular}
    \end{small}
\end{table*}

The observed targets are listed in Table~\ref{t:targets_sample}. The full target designations are given in the table in the form WISE Jhhmmss.ss$\pm$ddmmss.s (except for CFBDS J013302$+$023128). We abbreviate source names to the short form hhmm$\pm$ddmm hereafter. With reported spectral types $\geq$T8 and estimated masses $\lesssim$40 M$_\mathrm{Jup}$ (see Section~\ref{BD_masses}), these objects are some of the coolest and lowest-mass known brown dwarfs in the Solar neighbourhood. The sample includes 24 late-Ts and 9 Y dwarfs, making it the largest sample of Y dwarfs targeted for binarity to date.

\subsection{Observing Strategy}
\label{strategy}

All targets were observed with the Infrared (IR) channel of the Wide Field Camera 3 (WFC3) instrument on the \textit{Hubble Space Telescope}, as part of GO program 15201 (PI Fontanive). Each object was observed for a full orbit, split equally between the F127M and F139M filters, using a similar strategy to that from \textit{HST} SNAP 12873 and described in \citet{Fontanive2018}. The combination of these bandpasses exploits a water absorption feature seen in late-type objects at 1.4~$\mu$m which can be used to identify low-mass objects in the field of view based on F127M--F139M colours. Indeed, subtellar spectra exhibit a deep water band covered by the F139M filter and not seen in stars with spectral types early than M6, while the F127M filter covers the $J$-band peak of brown dwarfs. A considerable drop in flux between the two bandpasses can hence be used to robustly identify the targets and possible faint companions (see \citealp{Fontanive2018} for details).

A total of 4 exposures of $\sim$300~s were taken in each filter in MULTIACCUM mode, for a total exposure time of $\sim$1200~s in each band. The specific NSAMP and SAMP-SEQ instrument parameters were adapted based on the visibility time of each orbit, in order to maximise the integration time for each target, and are reported in Table~\ref{t:observations}. In each filter, the 4 individual images were acquired along a 4-point box pattern, using large spacings of 2--3\arcsec\,between dithered positions. Table~\ref{t:observations} provides a summary of the observations.

\begin{table*}
    \addtolength{\tabcolsep}{2pt}
    \renewcommand{\arraystretch}{1.2}
    \centering
    \caption{Summary of \textit{HST} observations from GO program 15201.}
    \begin{small}
    \begin{tabular}{l c c c c c}
    \hline\hline
        Target Name & Obs. Date & \multicolumn{2}{c}{WFC3/IR F127M} & \multicolumn{2}{c}{WFC3/IR F139M} \\ \cmidrule(lr){3-4} \cmidrule(lr){5-6}
         &  (UT) & NSAMP/SAMP-SEQ & t (s) & NSAMP/SAMP-SEQ & t (s) \\
    \hline
        WISE 0005$+$3737  & 2018 Nov 17 & 13/SPAR25 & 1211.754 & 13/SPAR25 & 1211.754 \\
        WISE 0015$-$4615  & 2018 Mar 31 & 13/SPAR25 & 1211.754 & 13/SPAR25 & 1211.754 \\
        WISE 0032$-$4946  & 2018 May 25 & 13/SPAR25 & 1211.754 & 13/SPAR25 & 1211.754 \\
        WISE 0038$+$2758  & 2017 Oct 25 & 11/STEP50 & 1196.929 & 11/STEP50 & 1196.929 \\
        WISE 0049$+$2151  & 2017 Oct 26 & 11/STEP50 & 1196.929 & 11/STEP50 & 1196.929 \\
        CFBDS 0133$+$0231  & 2018 Feb 10 & 11/STEP50 & 1196.929 & 11/STEP50 & 1196.929 \\
        WISE 0325$-$3854  & 2018 May 13 & 13/SPAR25 & 1211.754 & 13/SPAR25 & 1211.754 \\
        WISE 0325$-$5044  & 2018 Mar 23 & 13/SPAR25 & 1211.754 & 13/SPAR25 & 1211.754 \\
        WISE 0350$-$5658  & 2018 Aug 27 & 13/SPAR25 & 1311.756 & 13/SPAR25 & 1311.756 \\
        WISE 0359$-$5401  & 2019 Jan 27 & 13/SPAR25 & 1211.754 & 13/SPAR25 & 1211.754 \\
        WISE 0404$-$6420  & 2018 Aug 24 & 13/SPAR25 & 1311.756 & 13/SPAR25 & 1311.756 \\
        WISE 0410$+$1502  & 2017 Dec 23 & 11/STEP50 & 1196.929 & 11/STEP50 & 1196.929 \\
        WISE 0413$-$4750  & 2018 May 27 & 13/SPAR25 & 1211.754 & 13/SPAR25 & 1211.754 \\
        WISE 0744$+$5628  & 2019 Oct 06 & 14/SPAR25 & 1311.756 & 14/SPAR25 & 1311.756 \\
        WISE 0811$-$8051  & 2017 Nov 25 & 13/SPAR25 & 1311.756 & 13/SPAR25 & 1311.756 \\
        WISE 0812$+$4021  & 2017 Nov 05 & 13/SPAR25 & 1211.754 & 13/SPAR25 & 1211.754 \\
        WISE 0857$+$5604  & 2018 Nov 09 & 14/SPAR25 & 1311.756 & 14/SPAR25 & 1311.756 \\
        WISE 0943$+$3607  & 2018 Feb 11 & 13/SPAR25 & 1211.754 & 13/SPAR25 & 1211.754 \\
        WISE 1051$-$2138  & 2018 Jun 25 & 11/STEP50 & 1196.929 & 11/STEP50 & 1196.929 \\
        WISE 1206$+$8401  & 2018 Apr 01 & 14/SPAR25 & 1311.756 & 14/SPAR25 & 1311.756 \\
        WISE 1318$-$1758  & 2018 Jun 21 & 11/STEP50 & 1196.929 & 11/STEP50 & 1196.929 \\
        WISE 1541$-$2250  & 2018 Feb 17 & 11/STEP50 & 1196.929 & 11/STEP50 & 1196.929 \\
        WISE 1639$-$6847  & 2019 Mar 11 & 14/SPAR25 & 1311.756 & 14/SPAR25 & 1311.756 \\
        WISE 1738$+$2732  & 2017 Nov 07 & 11/STEP50 & 1196.929 & 11/STEP50 & 1196.929 \\
        WISE 2019$-$1148  & 2017 Oct 26 & 11/STEP50 & 1196.929 & 11/STEP50 & 1196.929 \\
        WISE 2056$+$1459  & 2018 Sep 12 & 11/STEP50 & 1196.929 & 11/STEP50 & 1196.929 \\
        WISE 2102$-$4429  & 2018 Apr 15 & 13/SPAR25 & 1211.754 & 13/SPAR25 & 1211.754 \\
        WISE 2212$-$6931  & 2018 Aug 27 & 14/SPAR25 & 1311.756 & 14/SPAR25 & 1311.756 \\
        WISE 2255$-$3118  & 2017 Oct 25 & 13/SPAR25 & 1211.754 & 13/SPAR25 & 1211.754 \\
        WISE 2313$-$8037  & 2018 Mar 25 & 14/SPAR25 & 1311.756 & 14/SPAR25 & 1311.756 \\
        WISE 2325$-$4105  & 2018 May 30 & 13/SPAR25 & 1211.754 & 13/SPAR25 & 1211.754 \\
        WISE 2332$-$4325  & 2017 Oct 25 & 13/SPAR25 & 1211.754 & 13/SPAR25 & 1211.754 \\
        WISE 2354$+$0240  & 2018 Jun 18 & 11/STEP50 & 1196.929 & 11/STEP50 & 1196.929 \\
    \hline \\ [-2.5ex]
    %
    \label{t:observations}
    \end{tabular}
    \end{small}
\end{table*}

\subsection{Data Reduction}
\label{data_reduction}

In our reduction procedures, we used the \texttt{\_flt} images exclusively, which are corrected 
via standard calibrations (bias, dark and flat field), provide fluxes in units of electron per second, 
but preserve the pixel data with their original sampling, a fundamental characteristic for careful 
stellar profile fitting in under-sampled images \citep{AndersonKing2000}.
We measured positions and fluxes in each of these images using 
the \texttt{hst1pass} code, which is a generalisation of the software \texttt{img2xym\_WFC}, 
initially developed to perform Point Spread Function (PSF) fitting in the Wide Field Channel (WFC) 
of the Advanced Camera for Surveys (ACS) of \textit{HST} \citep{AndersonKing2006}. 
The code perturbs a library PSF in order to empirically find the best spatially variable PSF for each image. 
With this PSF, \texttt{hst1pass} then runs a single pass of source findings without performing neighbour subtraction, so as to obtain 
initial positions and fluxes. Positions and fluxes are then corrected for the geometric distortion \citep{Anderson2016}. 
The library PSFs and the WFC3/IR geometric distortion developed by J.\ Anderson are both publicly 
available.\footnote{\url{https://www.stsci.edu/~jayander/STDPSFs/} and\\ \url{https://www.stsci.edu/~jayander/STDGDCs/}}

For each of our targets, one of these single-exposure catalogues in F127M was adopted as the common pixel-based reference coordinate system for the particular target's field. 
We then used well-measured unsaturated stars to derive general six-parameter linear transformations to transform stellar positions 
from all other images into the common reference frame for the selected target. 
Similarly, magnitudes were zero-pointed to the adopted common reference frame in each filter.

\subsection{Source Detection and Photometry}
\label{source_detection}

Once this preliminary photometry had provided all the transformations from 
all exposures into a common coordinate and photometric reference system, we ran a more 
sophisticated computer program to obtain photometry, as well as Artificial Star Tests (ASTs; see Section~\ref{ASTs}). 
This software package, \texttt{KS2}, developed by J.\, Anderson \citep{Anderson2008}
uses the previously obtained transformations and PSFs to simultaneously find and measure 
stars in all of the individual exposures, across the full fields of view of the images, and for both filters. 
Combining the multiple exposures, \texttt{KS2} finds and measures those faint stars that would 
otherwise be lost in the noise of single images. 
Detailed descriptions and examples of the usage of \texttt{KS2} are given in \citet{Bellini2017} and \citet{Nardiello2018}, and in 
a more recent application by \citet{Scalco2021}.

Given the sparse nature of the studied fields, which gave less control on the exact PSF shapes, 
we employed a single wave of finding, therefore limiting the search to sources separated by more than a pixel from each other.
Along with photometry and astrometry, \texttt{KS2} also provides important diagnostic parameters, such as: 
the quality-fit ("QFIT") parameter, which gives a measure of how accurate the PSF fit is, the photometric root mean square ("rms") for 
single exposure measurements, the contamination parameter from neighbours ("o"), the local sky value ("SKY") and its root mean square ("rmsSKY"), and 
the "RADXS" parameter (defined as in \citealp{Bedin2008}) which quantifies how much the flux distribution of a source represents that of the 
PSF. Extended sources have large positive RADXS values, hot pixels and cosmic rays have large negative values, and point-sources have values close to 0. 
%
With detailed transformation in positions and magnitudes, we also created stacked images of the astronomical scene (one per filter) 
super-sampled by a factor 2 (as described in \citealp{Scalco2021}). 
Photometry was zero-pointed to the Vega-magnitude system following the 
recipes in \citet{Bedin2005}, using encircled energy and ZP available 
in the STScI webpages for WFC3/IR\footnote{
\url{https://www.stsci.edu/hst/instrumentation/wfc3/data-analysis/photometric-calibration}
}.
The adopted zero-points to transform from instrumental magnitudes into calibrated magnitudes are 
23.55 and 23.25 for the F127M and F139M filters, respectively.

\subsection{Artificial Star Tests}
\label{ASTs}

Artificial star tests (ASTs) have a major role in most imaging investigations, enabling assessment of the accuracy of input and output photometry, and determining potential systematic bias functions in positions or stellar magnitudes.
For our ASTs, we added to individual images artificial stars as described in \citet{Anderson2008}, 
using our knowledge of PSFs, photometric, and astrometric transformations described in the previous section.
Each artificial star was randomly attributed a position, and F127M and F139M fluxes based on adopted distributions described below, and injected accordingly in the respective datasets in each band. The software adds and measures the artificial stars one at a time, therefore not creating a fake overcrowding. 

For each target, we added $4\times10^5$ artificial stars uniformly distributed across the full images, and with flat distributions in the colour-magnitude plane using instrumental magnitudes randomly drawn between $-5$ and $+6$ in both F127M and F139M (corresponding to 18.55--29.55 in $m_{\rm F127M}$ and 18.25--29.25 in $m_{\rm F139M}$). 
By comparing the positions and magnitudes of the injected and retrieved stars, the ASTs can be used to select adopted thresholds for the search of companions (Section~\ref{comp_search}), as well as to determine the completeness of the survey (Section~\ref{completeness}).
We explored the input and recovered positions and magnitudes of the artificial stars, together with the various measured quality factors (QFIT, RADXS), in order to determine the thresholds to be used in our source selections. 

We started by considering as robustly detected all sources with consistent detections in at least 3 of the 4 images in a given filter, and with a measured flux larger than the local rmsSKY value. All other detections were discarded for being dubious, below the thresholds adopted for a trustworthy detection. We then considered to be well recovered the artificial stars with a retrieved position within 1 pixel from the inserted one in both the X and Y directions, and with a measured magnitude $< -2.5 \times \log_{10}(2)$ from the injected one in a given filter. All sources not satisfying these criteria were counted as being poorly measured. From this, we examined the properties of well and poorly recovered sources in the obtained ASTs catalogues, to define criteria that would enable us to select only well-measured sources in the original datasets, while also excluding as few as possible. After exploring the numerous variables available and various combinations of the parameter space for the well and poorly-measured sources, we arrived at the following sequence: 
\begin{itemize}
    \item ratio of flux from contaminating neighbours to that of the measure source "o" $<$ 4
    \item quality of fit QFIT $>$ 0.9
    \item $-0.35 < \mathrm{RADXS} < 0.35$
\end{itemize}

\begin{figure*}
    \centering
    \includegraphics[width=\linewidth]{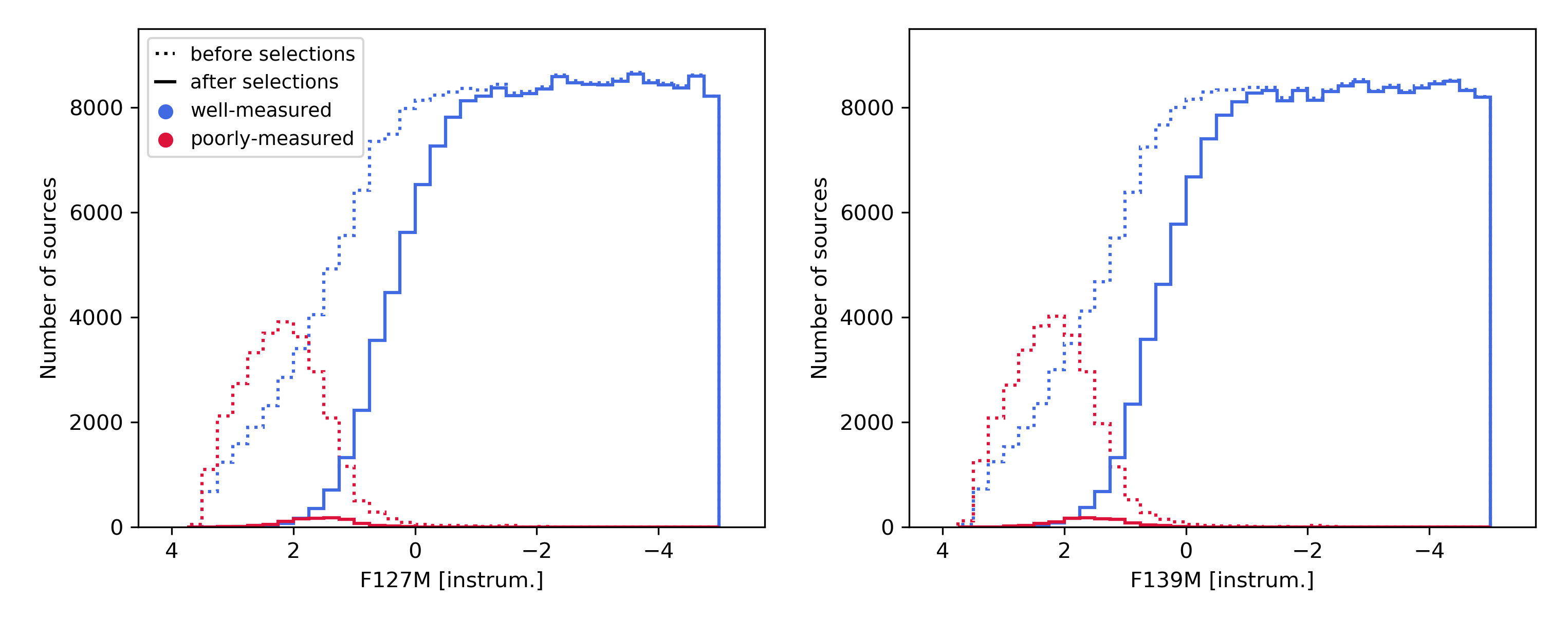}
    \caption{Magnitudes distributions in F127M (left) and F139M (right) of all injected stars with consistent detections in $\geq$3/4 images and measured fluxes above the  local sky value (dotted lines), compared to those left after applying the selection criteria described in the text (solid line). The blue and red histograms represent artificial stars that were well and poorly measured by the algorithms, respectively, showing that our adopted selections successfully exclude the vast majority of poorly-measured sources. This example is shown for the analyses made on the data for WISE~1639$-$6847.}
    \label{f:ASTs_hist}
\end{figure*}

\begin{figure*}
    \centering
    \includegraphics[width=\linewidth]{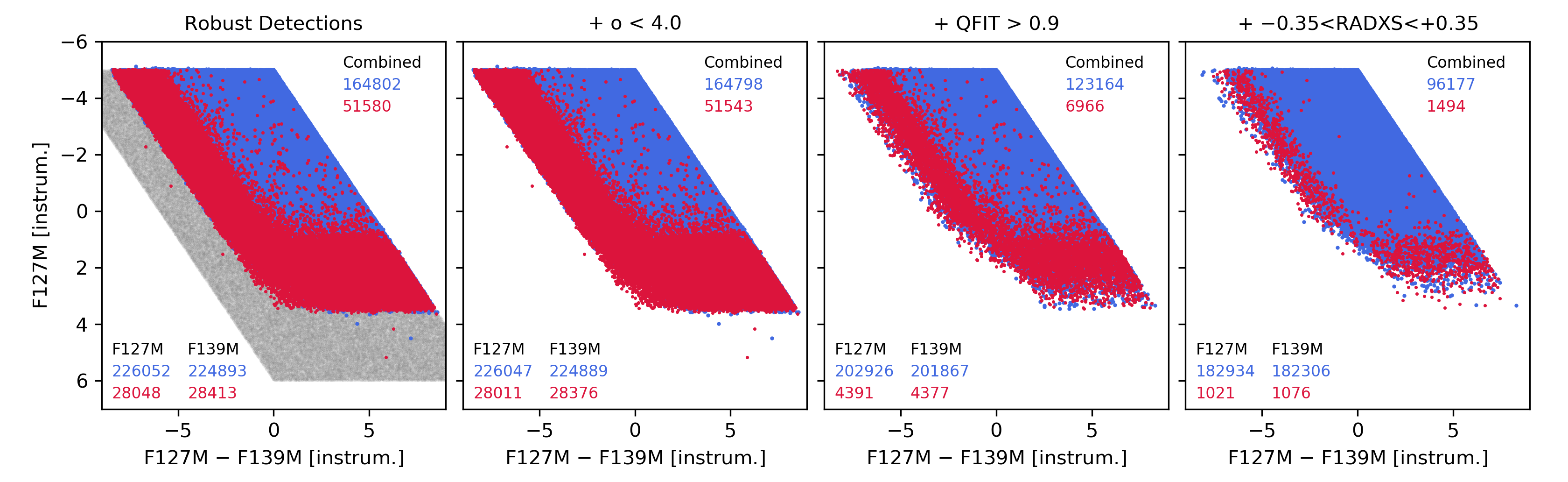}
    \caption{Injected artificial stars in the F127M--F139M colour-magnitude diagram. Each panel shows, from the left to the right, the well (blue) and poorly (red) measured stars that are left after each step of the selection process established from our analyses. The first panel shows the full set of simulated stars, where sources that did not satisfy the initial criteria for a detection are plotted in grey.}
    \label{f:ASTs_CMD}
\end{figure*}

Figures~\ref{f:ASTs_hist} and Figure~\ref{f:ASTs_CMD} show the performance reached by applying these criteria in the source selection process, for the example target WISE~1639$-$6847, chosen because the same datasets had already been extensively studied by our team in \citealp{BedinFontanive2020} and \citealp{Fontanive2021}. In Figure~\ref{f:ASTs_hist}, we show the distributions in measured F127M and F139M instrumental magnitudes of the well (blue) and poorly (red) measured artificial stars, before (dotted histograms) and after (solid histograms) applying the selection criteria.
The four panels in Figure~\ref{f:ASTs_CMD} show each step of the process, in the F127M--F139M vs. F127M colour-magnitude space. The numbers of well-recovered (blue) and badly-measured (red) sources remaining after each step are written in the top right corner on each panel. However, only artificial stars detected in both filters can be included in the plot, and sources were plotted in blue when they were found to have been successfully recovered in \textit{both} bands, and in red when the retrieved magnitudes or positions were too deviant from the injected ones in \textit{at least one} bandpass. This means that the plots show a larger number of poorly-recovered sources relative to the well-measured ones, compared to the same fraction within the separate samples from each band. We therefore also indicate in each panel the numbers of well and poorly measured artificial stars in the F127M and F139M data independently, which, for the first and last panels, correspond to the respective numbers of sources in the histograms from Figures~\ref{f:ASTs_hist}. In the first panel, the full set of injected artificial stars is also shown, with sources that were not detected at all, or did not meet the initial selection for a robust detection, shown in grey (plotted using the injected magnitudes for lack of recovered values in many cases). 

Across all targets, the final criteria used provide a recoverability rate for well-measured sources of $\sim$80--85\% in both filters, over the magnitude ranges explored in the ASTs. Most of the lost sources are at the faint end of the robustly-detected subset, at instrumental magnitudes $>$1--2, with $>$90\% of well-measured sources brighter than $m_\mathrm{instr.}$ of $+$1, and $>$99\% of those brighter than $m_\mathrm{instr.}$ of $-$1, retained in either filter after applying the above criteria. These effects can be rigorously quantified and accounted for with a completeness analysis, as performed in Section~\ref{completeness}. Most importantly, the final fraction of badly-measured sources that were not excluded with the adopted criteria was found to be $<$4--5\%, and again to mostly be for sources fainter than $m_\mathrm{instr.}$ of 1 in a given filter, with similar results achieved for all targets. We therefore consider this chain of selection thresholds to successfully achieve the sought goals of identifying most well-measured stars, while rejecting the vast majority of badly-recovered sources.

\section{Data Analyses of the Brown Dwarf Primaries}
\label{data_analysis:primaries}

\subsection{Photometric Properties}
\label{BD_photometry}

For each system, the primary brown dwarf target was identified in the list of the detected sources from the procedures described in Section~\ref{source_detection}, and its Vega magnitudes extracted from the resulting instrument photometry and corresponding zeropoints (ZP$_\mathrm{F127M}$ = 23.55, ZP$_\mathrm{F139M}$ = 23.25). As expected from their late spectral types, all probed objects showed a deep drop in the F139M water-band filter, due to the 1.4-$\mu$m water absorption feature seen in brown dwarfs. When the absorption was so strong that the target dropped below the detection level in the F139M images, an upper limit on the F139M flux of the brown dwarf was derived as a 5-$\sigma$ threshold above the local sky background, using the local noise values derived in the procedures from Section~\ref{source_detection}.
The measured \textit{HST} photometry for all targets is reported in Table~\ref{t:targets_properties}, along with \textit{Spitzer} photometry from \citet{Kirkpatrick2019}.

\begin{table*}
    \addtolength{\tabcolsep}{2pt}
    \renewcommand{\arraystretch}{1.2}
    \centering
    \caption{Apparent photometry and derived properties of the studied targets.}
    \begin{small}
    \begin{tabular}{l c c c c c c c}
    \hline\hline
        Target Name & WFC3/F127M & WFC3/F139M & IRAC/$ch1$   & IRAC/$ch2$   & $T_\mathrm{eff}$ & Luminosity & Mass \\
             & (mag) & (mag) & (mag) & (mag) & (K)  & log($L/L_\odot$) & (M$_{\mathrm{Jup}}$)\\
    \hline
        WISE 0005$+$3737  & $16.89\pm0.12$ & $23.26\pm0.13$ & $15.431\pm0.024$ & $13.282\pm0.018$ & $557\pm25$ & $-6.13\pm0.05$ & $27\pm7$ \\
        WISE 0015$-$4615  & $17.17\pm0.01$ & $23.27\pm0.47$ & $16.096\pm0.031$ & $14.228\pm0.019$ & $626\pm29$ & $-5.95\pm0.05$ & $32\pm8$ \\
        WISE 0032$-$4946  & $17.88\pm0.01$ & $24.44\pm0.70$ & $16.932\pm0.049$ & $14.929\pm0.021$ & $584\pm34$ & $-6.06\pm0.08$ & $29\pm8$ \\
        WISE 0038$+$2758  & $17.93\pm0.01$ & $24.49\pm0.12$ & $16.454\pm0.037$ & $14.410\pm0.020$ & $514\pm19$ & $-6.26\pm0.04$ & $24\pm7$ \\
        WISE 0049$+$2151  & $15.73\pm0.05$ & $21.69\pm0.10$ & $15.009\pm0.022$ & $13.043\pm0.017$ & $611\pm42$ & $-6.00\pm0.10$ & $31\pm8$ \\
        CFDBS 0133$+$0231 & $17.56\pm0.01$ & $23.15\pm0.40$ & $16.789\pm0.044$ & $15.053\pm0.023$ & $596\pm92$ & $-6.05\pm0.24$ & $30\pm10$ \\
        WISE 0325$-$3854  & $18.19\pm0.03$ & $>24.89$ & $17.120\pm0.053$ & $14.984\pm0.021$ & $565\pm31$ & $-6.11\pm0.07$ & $28\pm7$ \\
        WISE 0325$-$5044  & $18.39\pm0.01$ & $24.24\pm0.53$ & $17.746\pm0.086$ & $15.696\pm0.025$ & $662\pm48$ & $-5.87\pm0.10$ & $34\pm9$ \\
        WISE 0350$-$5658  & $21.40\pm0.04$ & $>25.03$ & $17.936\pm0.096$ & $14.688\pm0.020$ & $320\pm12$ & $-7.00\pm0.06$ & $10\pm3$ \\
        WISE 0359$-$5401  & $20.95\pm0.01$ & $>24.93$ & $17.553\pm0.072$ & $15.326\pm0.023$ & $409\pm22$ & $-6.62\pm0.08$ & $17\pm5$ \\
        WISE 0404$-$6420  & $19.01\pm0.01$ & $>24.91$ & $17.633\pm0.082$ & $15.418\pm0.022$ & $580\pm36$ & $-6.07\pm0.08$ & $29\pm8$ \\
        WISE 0410$+$1502  & $18.51\pm0.01$ & $>24.81$ & $16.636\pm0.042$ & $14.166\pm0.019$ & $400\pm21$ & $-6.65\pm0.08$ & $16\pm5$ \\
        WISE 0413$-$4750  & $19.20\pm0.05$& $>24.81$ & $17.802\pm0.086$ & $15.487\pm0.024$ & $527\pm47$ & $-6.23\pm0.13$ & $25\pm7$ \\
        WISE 0744$+$5628  & $16.74\pm0.01$ & $22.28\pm0.02$ & $16.267\pm0.034$ & $14.552\pm0.020$ & $676\pm70$ & $-5.84\pm0.15$ & $35\pm10$ \\
        WISE 0811$-$8051  & $18.82\pm0.03$ & $>24.95$ & $16.817\pm0.045$ & $14.402\pm0.019$ & $459\pm25$ & $-6.44\pm0.07$ & $20\pm6$  \\
        WISE 0812$+$4021  & $17.38\pm0.01$ & $22.72\pm0.09$ & $16.932\pm0.048$ & $15.299\pm0.024$ & $848\pm63$ & $-5.48\pm0.10$ & $46\pm11$ \\
        WISE 0857$+$5604  & $16.78\pm0.02$ & $22.98\pm0.13$ & $16.018\pm0.030$ & $14.134\pm0.019$ & $611\pm45$ & $-5.99\pm0.10$ & $31\pm8$ \\
        WISE 0943$+$3607  & $19.10\pm0.01$ & $>24.87$ & $16.746\pm0.043$ & $14.284\pm0.019$ & $464\pm31$ & $-6.42\pm0.10$ & $21\pm6$ \\
        WISE 1051$-$2138  & $18.01\pm0.01$ & $24.09\pm0.31$ & $16.419\pm0.036$ & $14.598\pm0.020$ & $596\pm31$ & $-6.03\pm0.06$ & $30\pm8$ \\
        WISE 1206$+$8401  & $19.67\pm0.02$ & $>24.98$ & $17.339\pm0.061$ & $15.220\pm0.022$ & $414\pm14$ & $-6.60\pm0.05$ & $17\pm5$ \\
        WISE 1318$-$1758  & $17.70\pm0.01$ & $23.74\pm0.22$ & $16.790\pm0.045$ & $14.728\pm0.020$ & $600\pm30$ & $-6.02\pm0.06$ & $30\pm8$ \\
        WISE 1541$-$2250  & $20.46\pm0.04$ & $>24.74$ & $16.658\pm0.042$ & $14.228\pm0.019$ & $353\pm12$ & $-6.84\pm0.03$ & $13\pm4$ \\
        WISE 1639$-$6847  & $20.00\pm0.02$ & $>24.94$ & $16.186\pm0.018$ & $13.588\pm0.016$ & $361\pm14$ & $-6.82\pm0.05$ & $13\pm4$ \\
        WISE 1738$+$2732  & $18.88\pm0.02$ & $>24.95$ & $17.093\pm0.053$ & $14.473\pm0.019$ & $400\pm20$ & $-6.65\pm0.07$ & $16\pm5$ \\
        WISE 2019$-$1148  & $17.41\pm0.03$ & $23.34\pm0.07$ & $16.028\pm0.031$ & $14.251\pm0.020$ & $586\pm25$ & $-6.06\pm0.05$ & $29\pm8$ \\
        WISE 2056$+$1459  & $18.48\pm0.03$ & $>24.79$ & $16.031\pm0.030$ & $13.923\pm0.018$ & $427\pm10$ & $-6.55\pm0.04$ & $18\pm5$ \\
        WISE 2102$-$4429  & $17.65\pm0.02$ & $24.03\pm0.05$ & $16.325\pm0.036$ & $14.223\pm0.019$ & $527\pm20$ & $-6.22\pm0.05$ & $25\pm7$ \\
        WISE 2212$-$6931  & $19.07\pm0.04$ & $>25.07$ & $17.364\pm0.063$ & $14.973\pm0.021$ & $454\pm16$ & $-6.45\pm0.05$ & $20\pm6$ \\
        WISE 2255$-$3118  & $16.75\pm0.04$ & $22.15\pm0.06$ & $15.914\pm0.029$ & $14.210\pm0.019$ & $672\pm48$ & $-5.85\pm0.09$ & $35\pm9$ \\
        WISE 2313$-$8037  & $16.45\pm0.01$ & $21.65\pm0.16$ & $15.289\pm0.023$ & $13.683\pm0.018$ & $655\pm31$ & $-5.788\pm0.06$ & $34\pm9$ \\
        WISE 2325$-$4105  & $18.99\pm0.01$ & $>24.95$ & $16.264\pm0.033$ & $14.086\pm0.019$ & $458\pm31$ & $-6.44\pm0.10$ & $20\pm6$ \\
        WISE 2332$-$4325  & $18.64\pm0.02$ & $>24.76$ & $17.271\pm0.059$ & $15.012\pm0.021$ & $536\pm22$ & $-6.20\pm0.06$ & $26\pm7$ \\
        WISE 2354$+$0240  & $21.81\pm0.14$ & $>24.75$ & $18.105\pm0.109$ & $15.013\pm0.022$ & $335\pm11$ & $-6.93\pm0.04$ & $11\pm3$ \\ 
    \hline \\ [-2.5ex]
    \multicolumn{8}{l}{
    \begin{minipage}{0.92\textwidth}
        \textbf{Notes.} \textit{HST} photometry from this work, \textit{Spitzer} photometry from \citet{Kirkpatrick2019}.
        Effective temperatures, luminosities and masses are estimated with the \texttt{ATMO\,2020} evolutionary models from the \textit{HST}/F127M and \textit{Spitzer}/$ch2$ photometry, adopting ages of $5\pm3$~Gyr.
        
    \end{minipage}}
    \label{t:targets_properties}
    \end{tabular}
    \end{small}
\end{table*}

Of the 33 targeted brown dwarfs, 16 T8--T9 objects were successfully recovered in the F139M filter, thanks to the long total exposure times acquired ($\sim$1200--1300~s compared to $\sim$700~s in SNAP 12873). These detections provide measurements of F127M$-$F139M colours for the first time for such late type objects, allowing an extension to the known trend observed with spectral type \citep{Aberasturi2014,Fontanive2018} to later types. Only synthetic colours or minimum magnitude differences (from null detections in the F139M band) had been explored to date beyond mid-T spectral types. 

\begin{figure*}
    \centering
    \includegraphics[width=0.8\textwidth]{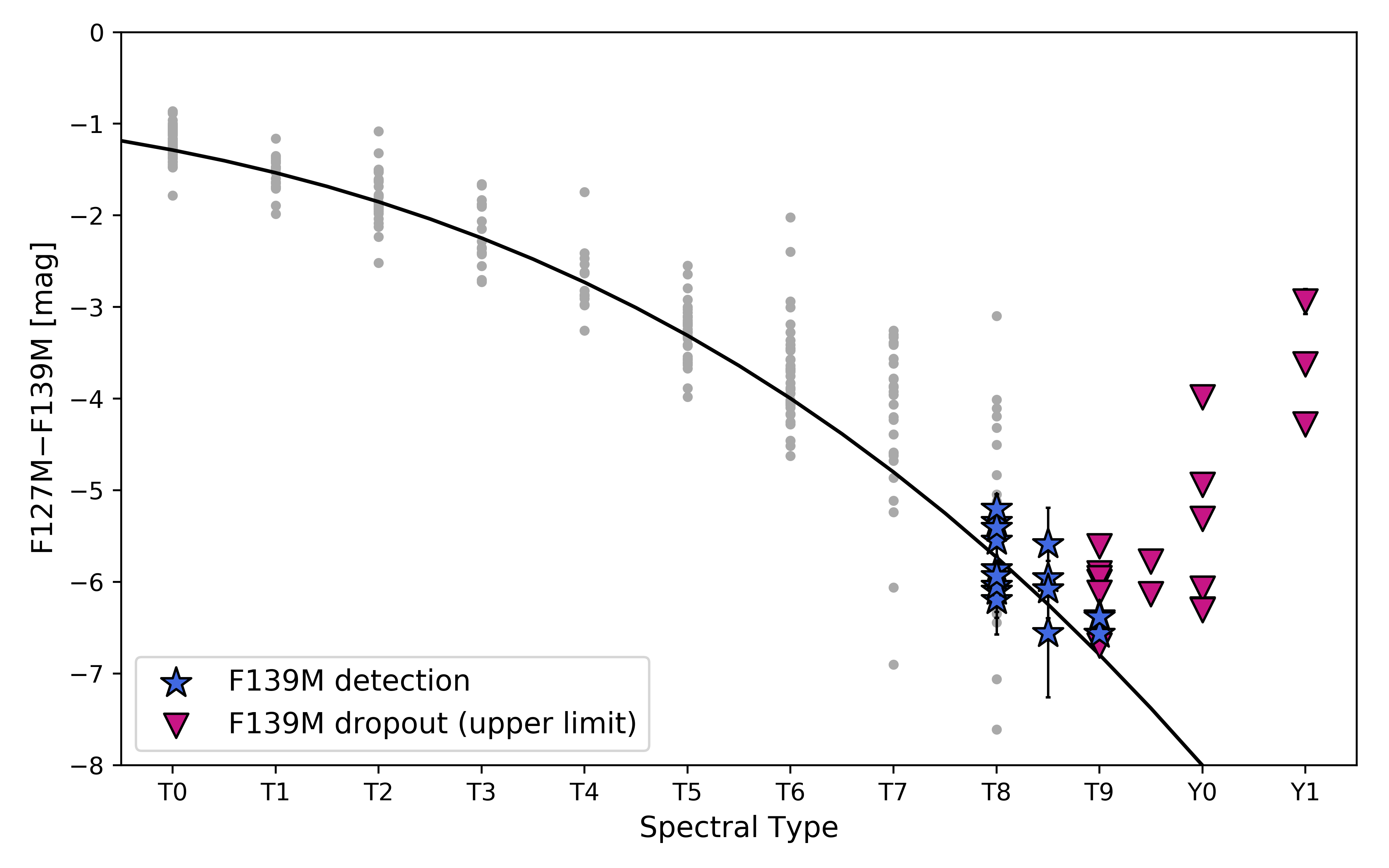}
    \caption{\textit{HST}/WFC3 F127M$-$F139M colours as a function of brown dwarf spectral type. Our new photometric measurements are displayed in the blue star symbols for targets recovered in the F139M band, and in the magenta triangles for water-band dropouts, for which only upper limits on the observed colours are available. The grey circles show synthetic colours for T-type sources in SPLAT. The solid black line corresponds to the polynomial fit from \citet{Fontanive2018}.}
    \label{f:SpT_Wabs}
\end{figure*}

Figure~\ref{f:SpT_Wabs} shows the \textit{HST} photometry as a function of spectral type for our targets with (blue stars) and without (magenta triangles) F139M detections. The newly measured colours for T8--T9 dwarfs are in good agreement with the extrapolation of the polynomial fit derived in \citet{Fontanive2018} for spectral types between T0 and T8, confirming that the water absorption band at 1.4~$\mu$m continues to deepen at the end of the T spectral sequence, with magnitude differences F127M$-$F139M of $-$5 to $-$6.5~mag for these systems, compared to values of about $-$3~mag for mid-Ts and around $-$1~mag for late Ls and early Ts \citep{Aberasturi2014,Fontanive2018}. For comparison, we also computed synthetic colours (grey circles) for sources from the SpeX Prism Library Analysis Toolkit (SPLAT; \citealp{SPLAT}), making use of the built-in functions in SPLAT to derive synthetic photometry from SpeX spectra. Synthetic colours for late-Ts show a much wider spread than our obtained measurements, most likely due to the noisy and low signal-to-noise nature of the available SpeX spectra for the faintest objects, in particular around the 1.4-$\mu$m water absorption band.

\begin{figure*}
    \centering
    \begin{subfigure}{0.3\textwidth}
        \centering
        \includegraphics[height=6.75cm]{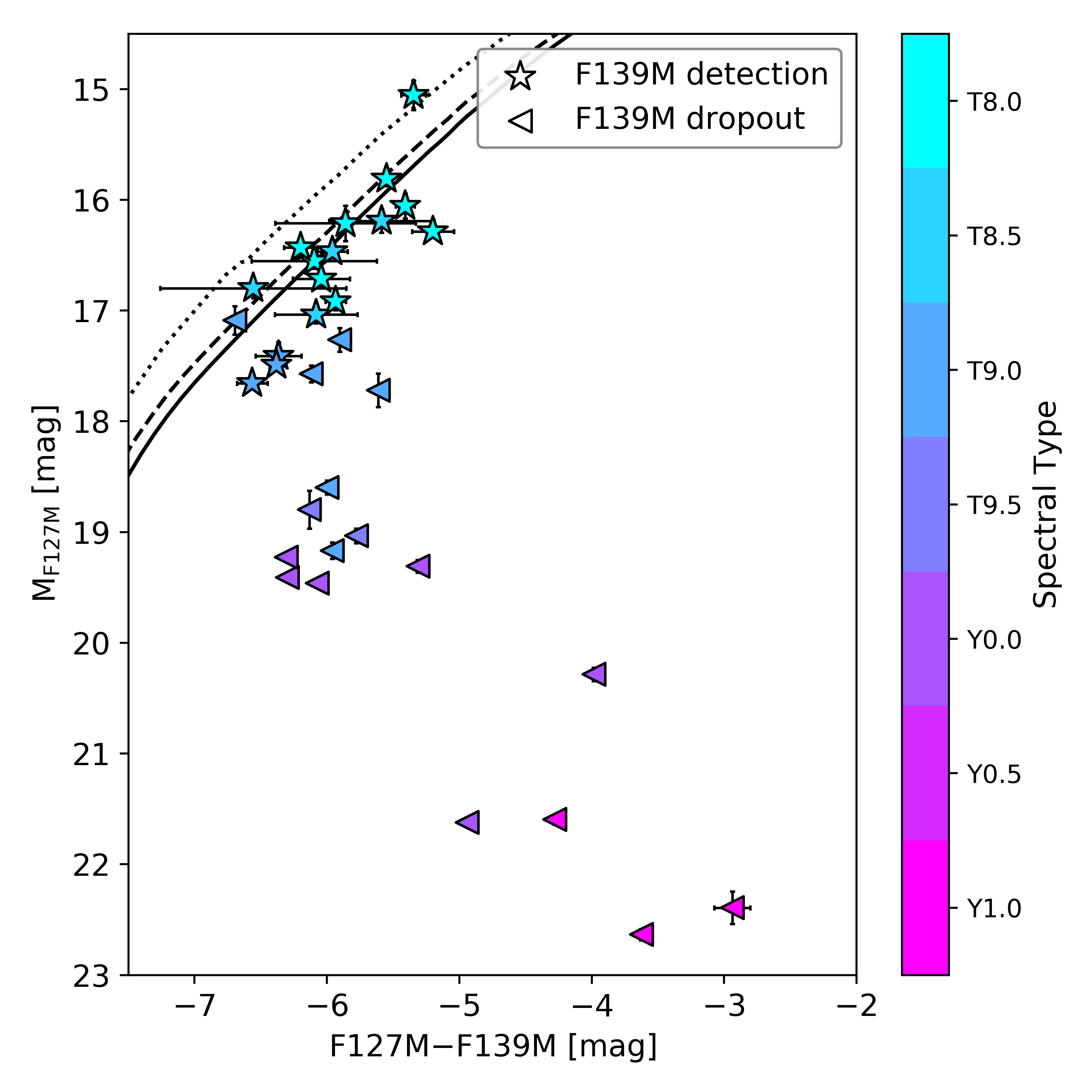}
    \end{subfigure}
    \hfill
    \begin{subfigure}{0.3\textwidth}
        \centering
        \includegraphics[height=6.75cm]{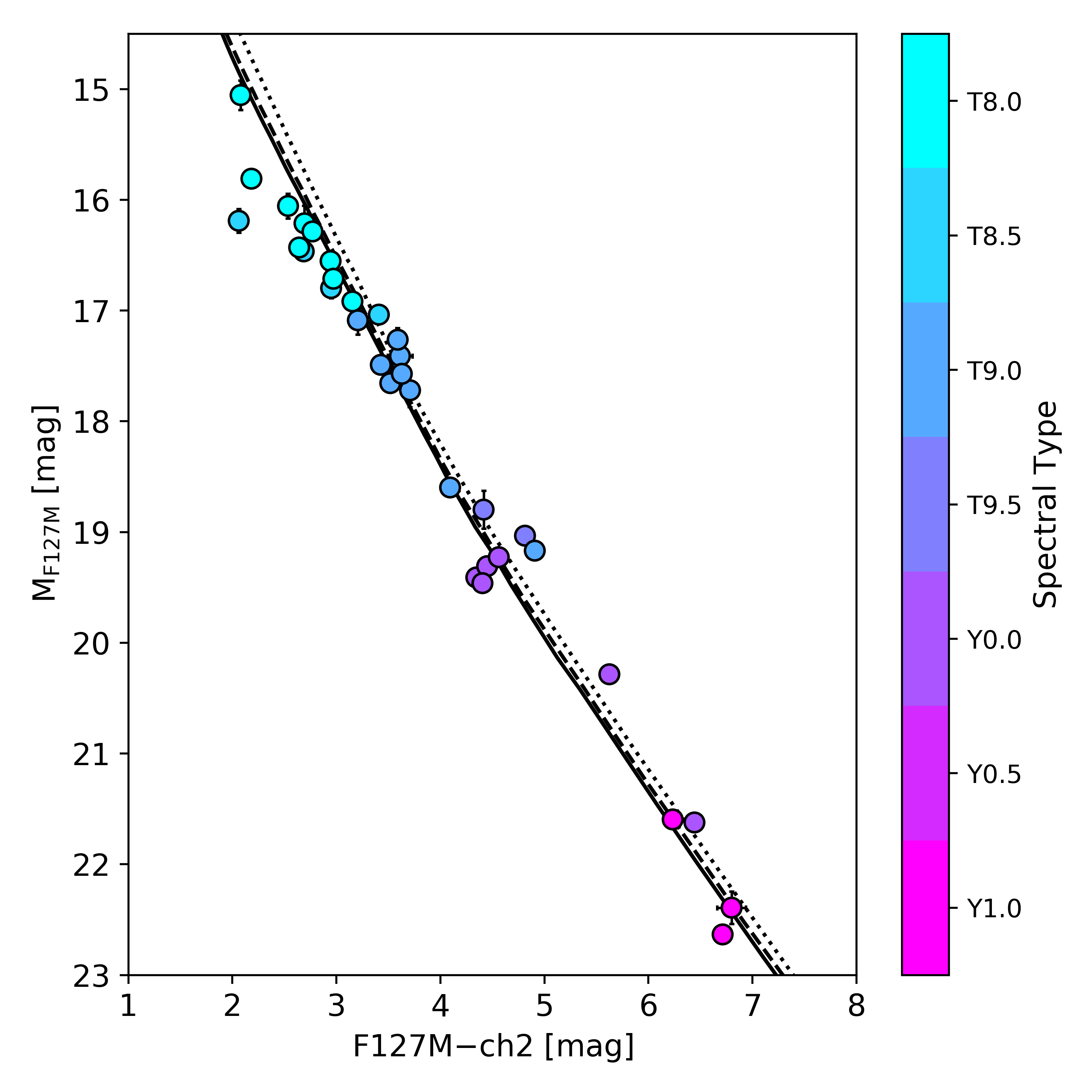}
    \end{subfigure}
    \hfill
    \begin{subfigure}{0.39\textwidth}
        \centering
        \includegraphics[height=6.75cm]{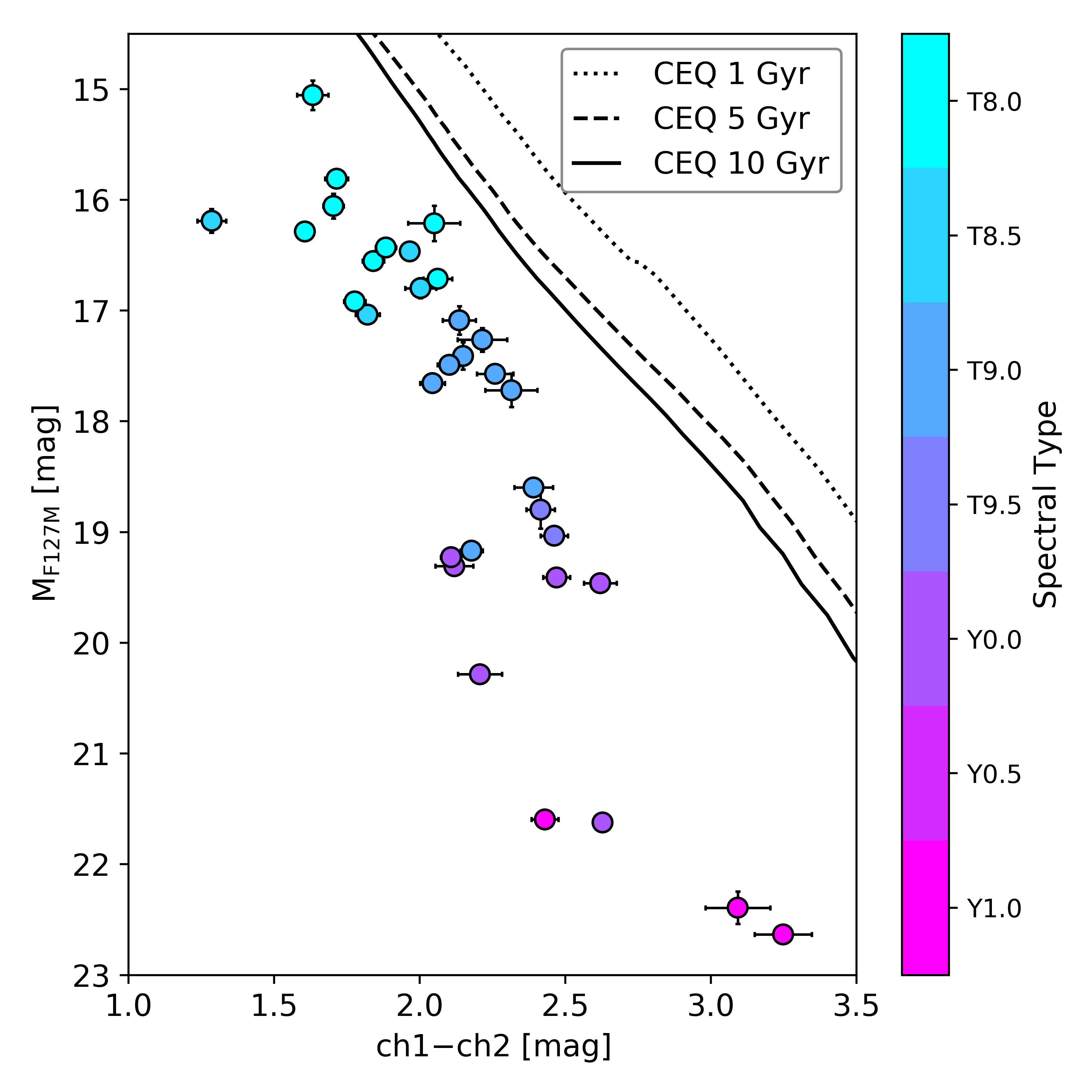}
    \end{subfigure}
    \caption{Colour-magnitude diagram of the 33 brown dwarf primaries, showing the \textit{HST}/WFC3 F127M absolute magnitudes versus their F127M$-$139M \textit{HST} colours (left), \textit{HST}$-$\textit{Spitzer} F127M--$ch2$ colours (centre), and infrared $ch1-ch2$ \textit{Spitzer} colours (right). 
    The colourbar indicates the spectral types of the objects. In the left panel, star symbols indicate sources successfully retrieved in the F139M band, thus plotting true colours. Triangles correspond to sources undetected in F139M for which the resulting colour value is therefore an upper limit (i.e., the true positions on the x-axis are to the left of the plotted points). In all three panels, the black lines correspond to the predicted colour-magnitude relationships from the \texttt{ATMO\,2020} models with equilibrium chemistry at discrete ages of 1~Gyr (dotted line), 5~Gyr (dashed line) and 10~Gyr (solid line). The poor fit of the models to the $ch1-ch2$ colours in the last panel is due to known issues of models around 4~$\mu$m, in $ch1$ band.}
    \label{f:Spitzer_HST_CMDs}
\end{figure*}

Figure~\ref{f:Spitzer_HST_CMDs} shows colour-magnitude diagrams of the observed targets, in terms of absolute F127M magnitudes against \textit{HST} (left), \textit{HST}$-$\textit{Spitzer} (middle) and \textit{Spitzer} (right) colours. The parallax measurements from Table~\ref{t:targets_sample} were used to convert apparent F127M magnitude into absolute photometry. Clear trends are observed in all cases with absolute brightness and spectral type, shown in the colourbar. Indeed, although no object $>$T9 was recovered in F139M (triangle symbols), the left panel demonstrates a continuous trend to bluer colours with fainter absolute F127M magnitude for T8 to T9 objects, as a result of the continuously deeper absorption level with colder effective temperatures in the 1.4-$\mu$m water-band feature (Figure~\ref{f:SpT_Wabs}). Similarly, the middle panel demonstrates a very tight relationship, with a continuous shift towards redder colours with dimmer F127M absolute flux and later spectral type (and thus cooler effective temperature). The right panel also exhibits a steady shift towards redder $ch1-ch2$ colours for fainter and later type objects, though not as tightly defined as for the \textit{HST}$-$\textit{Spitzer} colours.

The $ch1-ch2$ colours are also poorly matched by the models for given absolute F127M fluxes, despite the F127M--$ch2$ relationship agreeing very well with theoretical predictions. This is a known effect, sometimes referred to as the ``4-micron problem'' \citep{Leggett2019}, in which a deep CH$_4$ absorption feature in the 3--4~$\mu$m region is poorly reproduced by current models, which are consistently too faint in the L-band (and equivalent 3.6-$\mu$m \textit{Spitzer}/$ch1$ or \textit{WISE}/$W1$ bands) for objects cooler than $\sim$700~K (see also \citealp{Morley2018,Phillips2020,Leggett2021}).

\subsection{Primary Masses}
\label{BD_masses}

We used the \textit{HST} photometric data from this paper, together with the \textit{Spitzer} photometry of our targets from \citet{Kirkpatrick2019} and summarised in Table~\ref{t:targets_properties}, to estimate primary masses for our observed sample. Only the F127M fluxes were used from our \textit{HST} measurements, since half of our targets have no F139M detection. Furthermore, as available ground-based near-infrared photometry for our probed systems is highly heterogeneous in quantity, quality, spectral coverage, and filter system, we did not include additional sparse photometric data in this analysis. 

We adopted the \texttt{ATMO\,2020} models\footnote{\url{http://perso.ens-lyon.fr/isabelle.baraffe/ATMO2020/}. Models in the \textit{HST} filters were provided by M. Phillips via private communication.} \citep{Phillips2020}, a new set of coupled atmosphere and evolutionary grids for T and Y dwarfs with improved modelling and updated line lists. We used the predicted cooling tracks in the specific \textit{HST}/F127M and \textit{Spitzer}/$ch1$ and $ch2$ filters to estimate luminosities, effective temperatures, and age-dependent masses. Only the equilibrium chemistry (CEQ) models are considered here, although we achieved similar results on final inferred masses from the non-equilibrium models (CNEQ) with weak ($<$2~M$_\mathrm{Jup}$ differences) and strong ($<$5~M$_\mathrm{Jup}$ differences) vertical mixing (see \citealp{Phillips2020} for discussions of the various models and their fit to observations at different wavelengths), corresponding to $<$0.5-$\sigma$ disparities in all cases. Tracks from the CEQ sets of models are overplotted in Figure~\ref{f:Spitzer_HST_CMDs} at ages of 1, 5 and 10~Gyr. 

Absolute magnitudes in the F127M, $ch1$ and $ch2$ bandpasses were computed for all targets from the apparent photometry, and adopting trigonometric distances from the parallax measurements in Table~\ref{t:targets_sample}. We used a Monte-Carlo approach to propagate uncertainties, randomly drawing $10^5$ apparent magnitudes and parallaxes from Gaussian distributions centred on the measured values, with standard deviations set to the measurement uncertainties. Age estimates for isolated field brown dwarfs are particularly challenging to obtain, and the ages of our probed systems are unknown. We therefore assumed a similar distribution in age as in the solar neighbourhood \citep{Caloi1999}, and adopted a Gaussian distribution of ages centred around $5\pm3$~Gyr, as was done in \citet{Fontanive2018}. For each set of drawn parameters, the obtained absolute magnitudes were then interpolated in the \texttt{ATMO\,2020} evolutionary tracks at the generated system ages to infer corresponding luminosities, effective temperatures, and masses, thus providing 3 distributions for each inferred parameter, from the F127M, $ch1$, and $ch2$ data, respectively.

\begin{figure*}
    \centering
    \includegraphics[width=\textwidth]{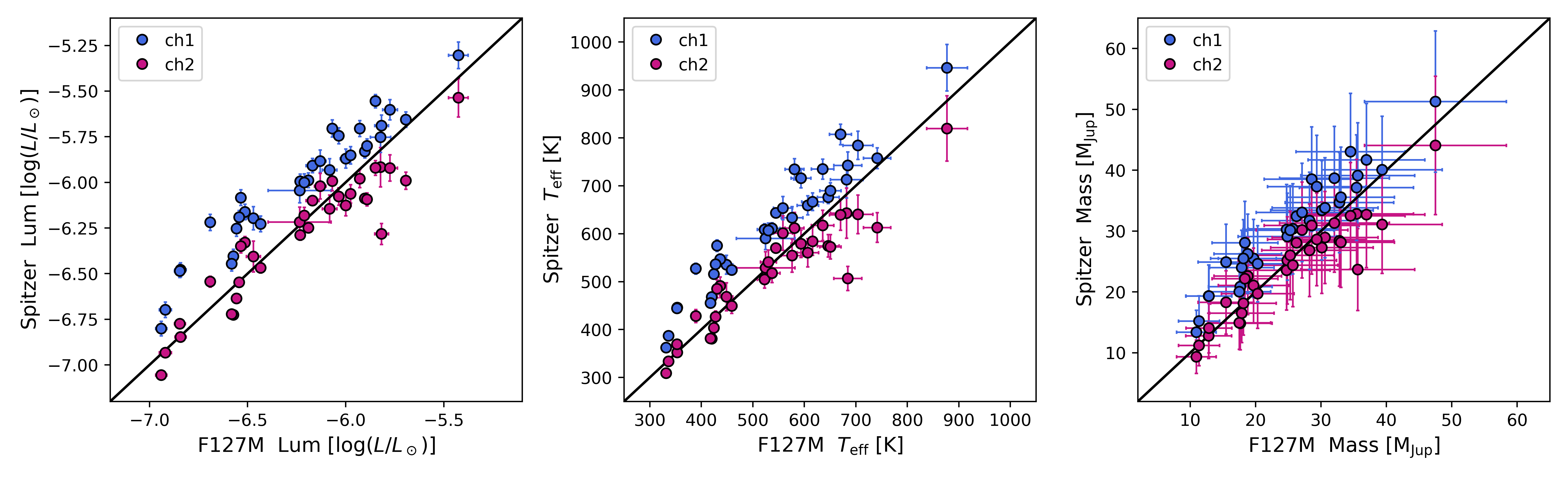}
    \caption{Model-derived parameters for our studied targets using the \texttt{ATMO\,2020} models \citep{Phillips2020}, comparing results obtained for the luminosity (left), effective temperature (centre) and mass (right) with the \textit{HST} F127M photometry to values derived from the \textit{Spitzer} $ch1$ (blue) and $ch2$ (magenta) photometry. The solid black lines in each panel show the 1:1 relationships. Results from the F127M and $ch2$ data are in good agreement, while values based on the $ch1$ photometry comparatively overestimate all three parameters. These differences become less significant in the right panel due to the strong age dependence of the derived masses, responsible for the much larger uncertainties.}
    \label{f:infered_params}
\end{figure*}

Figure~\ref{f:infered_params} shows comparisons of these individual distributions as the obtained F127M values against the $ch1$ (blue) and $ch2$ (magenta) values, plotted using the mean and standard deviation of each distribution from each target as the point value and its errorbar. The solid black lines mark the 1:1 relations. In general, we found the F127M and $ch2$ results to be rather consistent with each other, with mean offsets across the sample of 2.0 and 1.4~$\sigma$ in the obtained luminosities and effective temperatures, respectively, and a particularly good agreement for the faintest targets (log($L/L_\odot$) $<-6$; $T_\mathrm{eff}<700$~K; mass $<35$~M$_\mathrm{Jup}$). In comparison, parameters inferred from the $ch1$ photometry typically resulted in brighter luminosities, warmer effective temperatures, and larger masses, with mean discrepancies between the F127M and $ch1$ results around the 4-$\sigma$ level for both luminosities and effective temperatures. Offsets in the resulting masses were far less significant as a result of the age-dependence of the brown dwarf luminosity--mass relationship, responsible for the much wider uncertainties in the right-most panel. In contrast, luminosity and temperature are essentially age-independent properties for a given absolute magnitude, and the final errorbars in these parameters are dominated by the uncertainties in the apparent magnitudes and parallaxes measurements. The inferred masses based on \textit{Spitzer} photometry were found to agree with the F127M-derived masses at the 0.6-$\sigma$ level on average for $ch1$ estimates (1.2~$\sigma$ in the most discrepant case), and at 0.3~$\sigma$ on average for $ch2$ (1.0~$\sigma$ at most). The larger discrepancies seen with the $ch1$ data are consistent with the known issue of models in this band (see Section~\ref{BD_photometry}; \citealp{Phillips2020}).

Based on the recognised problems from the models in the 3--4~$\mu$m region, only the $ch2$ photometry was considered from \textit{Spitzer} to compute final values in each model-derived parameter. Final values of luminosity, effective temperature and mass are given in Table~\ref{t:targets_properties} for each target in the studied sample. These values and associated uncertainties  for each system correspond to the mean and standard deviation from the F127M and $ch2$ results simultaneously, i.e., combining the samples from the resulting distributions in each filter. Our estimated effective temperatures were found to be consistent with those from \citet{Kirkpatrick2019}, with $<$30~K differences (or agreeing within $<$1~$\sigma$) for the 31 targets in the sample (out of 33) that have estimated temperatures in \citet{Kirkpatrick2019}.
Offsets in the resulting masses were even less significant as a result of the large uncertainties induced by the strong dependence on age.

\section{Data Analyses for Multiplicity}
\label{data_analysis:binarity}

\subsection{Search for Companions}
\label{comp_search}

\subsubsection{Source Selection}

\begin{figure*}
    \centering
    \includegraphics[width=0.75\textwidth]{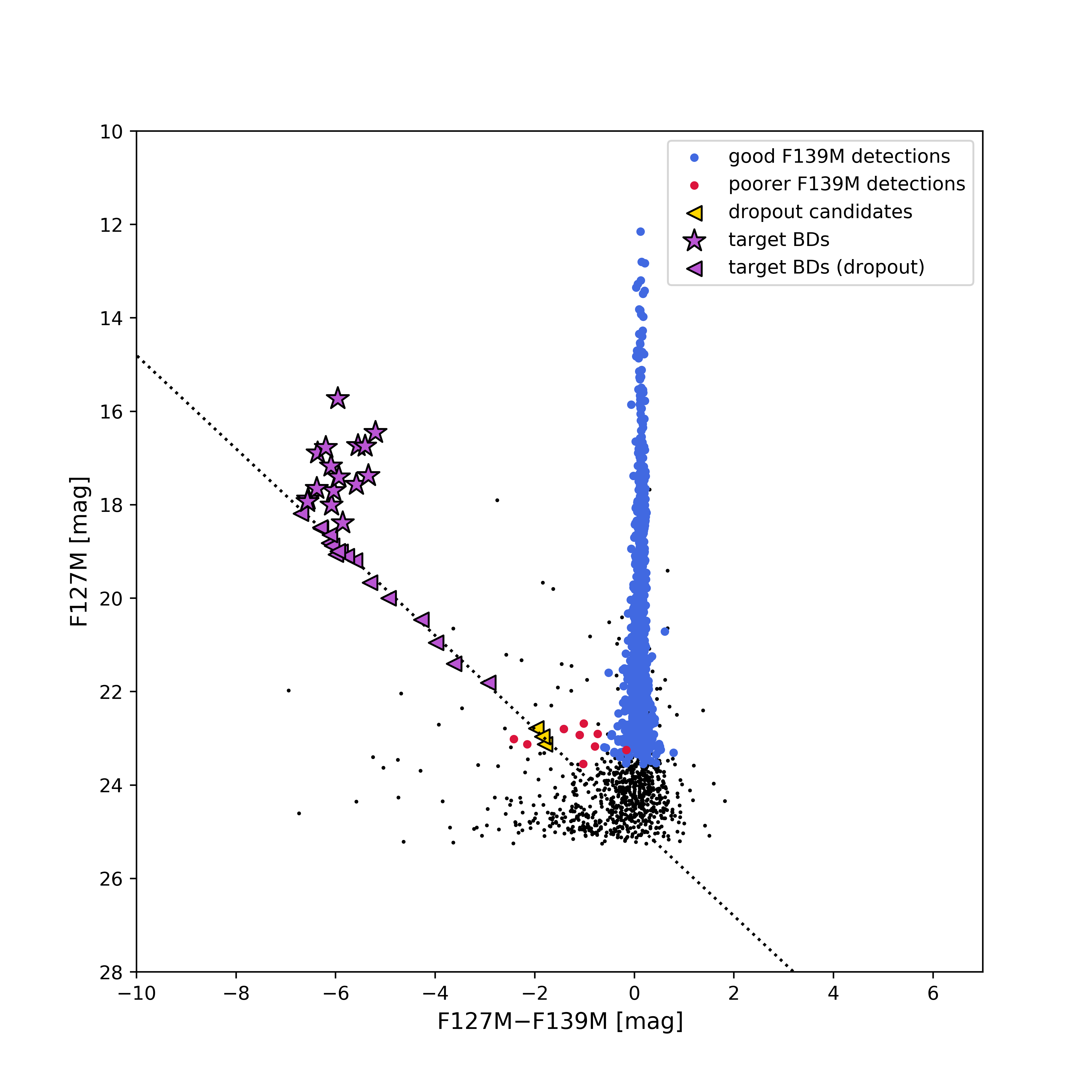}
    \caption{F127M--F139M colours vs. F127M apparent magnitudes for all sources well detected in F127M over all images of \textit{HST} program GO~15201, based on the criteria defined in Section~\ref{ASTs}. The black dots correspond to sources that did not satisfy the added selection of being 1.5~mag above the local F139M background value. Among sources satisfying this criterion, those well-measured in F139M are marked in blue, and those poorly measured in that band are shown in red. The brown dwarf targets are plotted in purple, distinguishing those recovered in F139M (star symbols) from those dropping out (triangles). The yellow triangles indicate the 3 sources identified as water-band dropout candidates.}
    \label{f:CMD_all_sources}
\end{figure*}

To search for resolved companions around our brown dwarf targets, we applied the source selection criteria defined in Section~\ref{ASTs}, from the results of the performed Artificial Star Tests. This provided us with catalogues of real sources from the F127M datasets, for which we can trust the reliability of measured fluxes and positions with a high confidence level.
Bonafide cold companions may be of two types: those recovered in the F139M filter, and those dropping out in the water-band. Any object from the first category must show a comparatively deep, or yet deeper, absorption level to the primary brown dwarfs, and would be easily identifiable in the colour-magnitude space, just like the primaries themselves (see Figure~\ref{f:CMD_all_sources}). Given that half of the science targets themselves were not detected in F139M, only companions as bright as our brighter targets would be recovered in F139M as well, since fainter objects with the colours of a cold brown dwarf rapidly fall below the detection level of the F139M bandpass, as is also the case for our fainter primaries. For such candidates, we therefore impose the criterion that sources must be fainter than the considered primary in F127M, with a threshold on the colour of the source of F127M--F139M $<$ [F127M--F139M]$_\mathrm{BD}$ $-$ 2, i.e., at most 2 magnitudes redder than the primary itself when the primary is retrieved in F139M.

Any dimmer potential companion would have to be a water-band dropout, detected in F127M but not in F139M, in order to be consistent with a true low-mass and cold substellar object. Such sources can be distinguished in the colour-magnitude diagrams (CMDs) when sufficiently bright, as outliers from the core of detected sources in the CMD, as is the case for the primaries in the sample. However, at fainter magnitudes, the minimum magnitude difference that may be measured for a dropout candidate (i.e. the upper limit on its F127M--F139M colour) approaches the blue end of the spread in F127M--F139M colours from other faint sources in the field, and thus cannot be confidently attributed to a water-band dropout candidate companion. Therefore, we imposed the additional constraints that selected candidates must be 1.5 magnitudes brighter in F127M than the local background magnitude in F139M. This ensured that retained candidates have maximum F127M--F139M colours bluer than the bulk of faint sources in the images, as it was found that the scatter in F127M--F139M colours for well-measured sources below that threshold approached the maximum colours measurable for a water-band dropout candidate (see Figure~\ref{f:CMD_all_sources}). 

Finally, any F139M detection for a source fainter than the primaries would make it highly improbable that that source is a true cold companion. We therefore also relaxed the constraints placed on the F139M source selections, since objects at the faint end may be missing from the F139M catalogues of well-measured sources (see Section~\ref{ASTs}), thus appearing as water-band dropouts, when in fact they should be disregarded as potential candidates for not being sufficiently blue. We hence removed the criteria placed on the F139M RADXS and QFIT parameters, and considered to also be recovered in F139M any source with detections in 3 out of the 4 individual F139M images, with a flux above the local rmsSKY.  

Figure~\ref{f:CMD_all_sources} shows the positions in the \textit{HST} colour-magnitude diagram of all well-detected F127M sources satisfying the above criteria, across all datasets from the survey, marked with different colours and symbols based on whether or not they were also detected in F139M. A similar figure was made and inspected for each target individually, but we display the combined sample of sources from all images in the same plot for simplicity, providing a summary of all results in the same context.
The brown dwarf primaries are shown as purple stars for those detected in the F139M filter, and plotted with purple triangles at the locations of their maximum F127M--F139M colours for those dropping out in F139M (their true positions are thus somewhere to the left of the plotted locations). The dotted black line shows where such dropout sources are expected to fall when plotted based on upper limits on colours, adopting an average F139M background magnitude value (5-$\sigma$). With a comparable sensitivity achieved across all datasets, all primaries undetected in F139M correctly fall along this line, and potential candidates dropping out in F139M are expected to do so as well.

\begin{table*}
    \addtolength{\tabcolsep}{2pt}
    \renewcommand{\arraystretch}{1.1}
    \centering
    \caption{Summary of identified candidate companions. For the photometry, candidates are marked as "cand", and the corresponding primaries as "BD".}
    \begin{small}
    \begin{tabular}{l c c c c c c c}
    \hline\hline
        Target & Sep. & Pos. Angle & F127M$_\mathrm{BD}$ & F127M$_\mathrm{cand}$ & F139M$_\mathrm{BD}$ & F139M$_\mathrm{cand}$ & Additional\\
         & (arcsec) & (deg) & (mag) & (mag) & (mag) & (mag) & HST data\\
    \hline
        WISE~0015$-$4615 & $30.41\pm0.02$ & $275.12\pm0.1$ & $17.17\pm0.01$ & $23.13\pm0.08$ & $23.27\pm0.47$ & $>24.92$ & GO~12972 \\
        WISE~1738$+$2732 & $25.72\pm0.02$ & $178.29\pm0.1$ & $18.88\pm0.02$ & $22.79\pm0.08$ & $>24.95$ & $>24.76$ & GO~12330, GO~16229\\
        WISE~2019$-$1148 & $35.82\pm0.02$ & $22.99\pm0.1$ & $17.41\pm0.03$ & $22.96\pm0.15$ & $23.34\pm0.07$ & $>24.81$ & ---\\
    \hline \\ [-2.5ex]
    %
    \label{t:candidates}
    \end{tabular}
    \end{small}
\end{table*}

\begin{figure}
    \centering
    \includegraphics[width=0.45\textwidth]{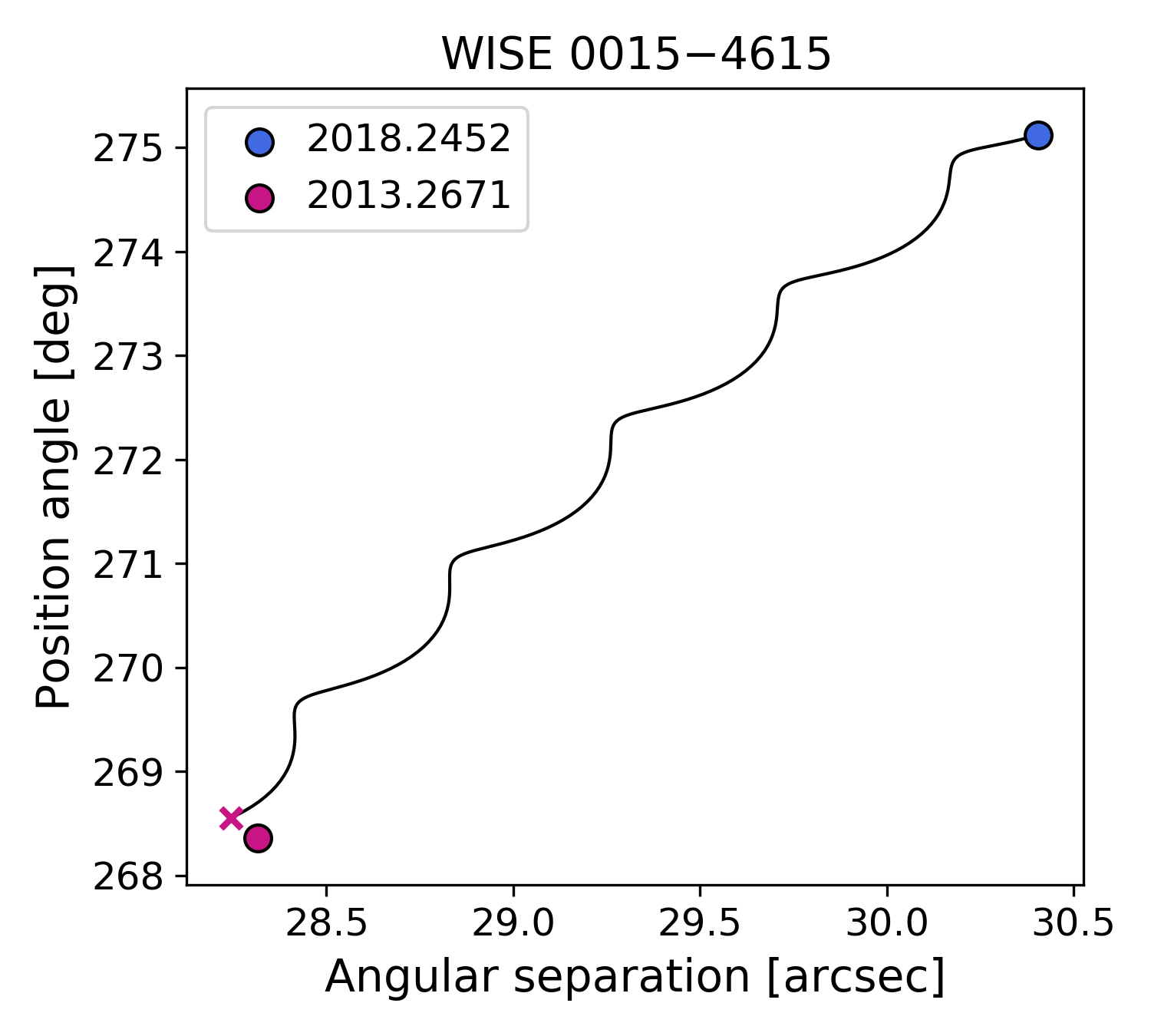}
    \caption{Common proper motion analysis of the candidate companion to WISE~0015$-$4615. The relative position of the candidate in the F127M images from our program is shown in the blue circle, and the black line shows the expected motion of a stationary background star relative to the primary between the available epochs. The relative position of the companion in archival \textit{HST} data from program GO~12972 (magenta circle) is consistent with the expected position at that epoch of a background source (magenta cross), confirming its background nature.}
    \label{f:W0015}
\end{figure}

\begin{figure}
    \centering
    \includegraphics[width=0.45\textwidth]{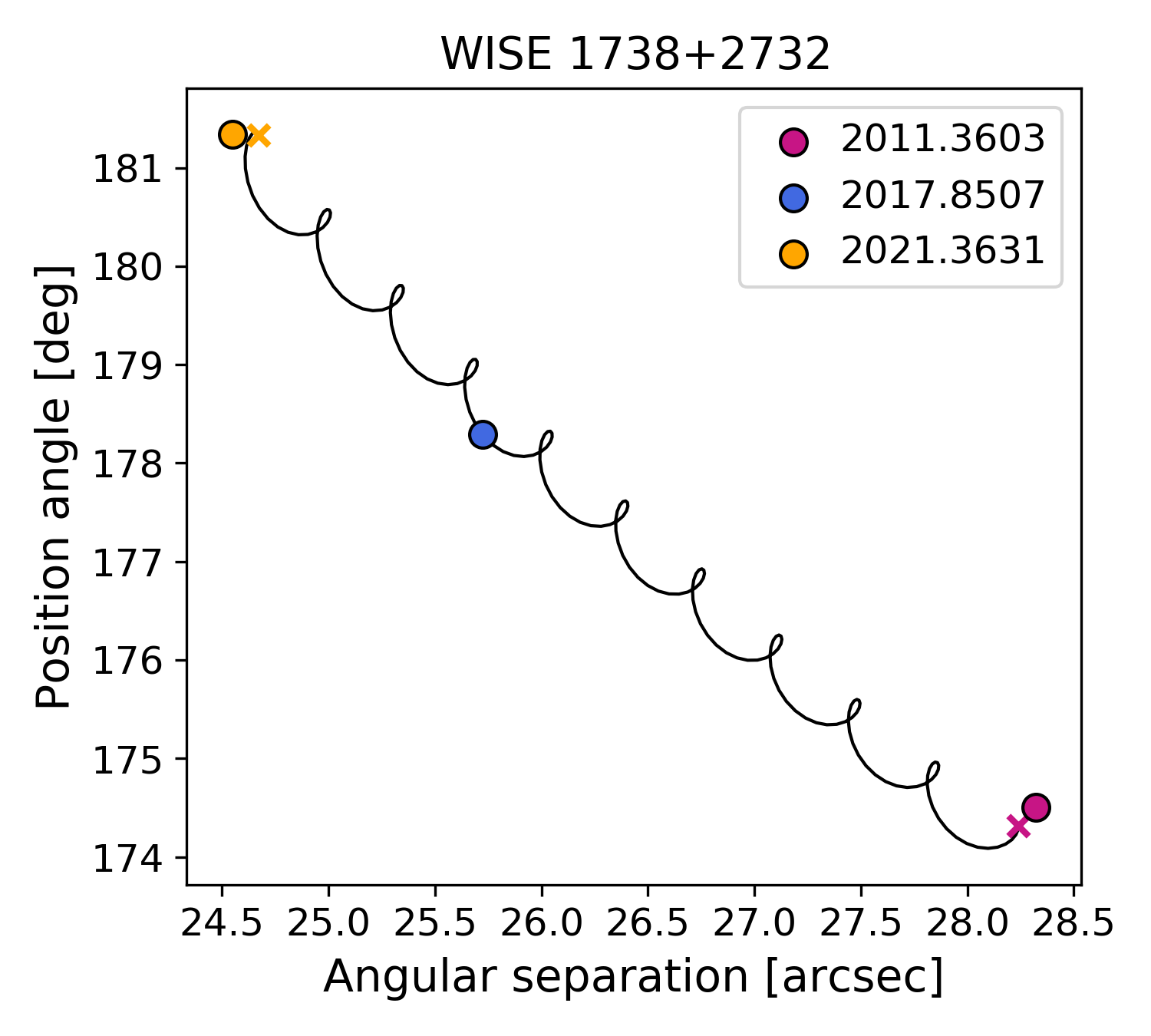}
    \caption{Common proper motion analysis of the candidate companion to WISE~1738$+$2732. The relative positions of the companion in archival \textit{HST} data from programs GO~12330 (magenta circle) and GO~16229 (yellow circle) are consistent with the expected position at those epochs of a background source (magenta and yellow crosses), confirming its background nature.}
    \label{f:W1738}
\end{figure}

No sources other than the brown dwarf targets themselves were found with very blue F127M--F139M colours, among those detected in F139M. Figure~\ref{f:CMD_all_sources} clearly shows the absence of such sources among those recovered in both filters, with a wide horizontal gap throughout all images between the primaries with F139M detections (purple stars) and other similarly-bright sources in F127M. No such blue candidate was identified upon visual examinations of the reduced data either, while the science targets were easily spotted in the final images. The handful of red scatter points indicate the few sources that did not satisfy the cuts for what was determined to be well-measured in F139M, but were nonetheless detected in that band, implying that they could be ruled out as possible dropout companions. Each of these sources was inspected individually and their F139M detections were visually confirmed in all cases, although determined to be poorly measured by the algorithms. The three yellow triangles are the only sources that remained as potential candidates, with faint F127M magnitudes and no F139M detections, and are discussed below. Finally, we found that bright but poorly measured F127M sources with rather blue colours (black dots above the dotted line and to the left of the core sample of detected sources, in blue) were mostly found to be non-real sources in the PSFs of bright stars, or to be galaxies, confirming that both categories of objects were successfully rejected by our source selections.

\subsubsection{Rejection of Candidate Companions}

The 3 identified candidates with significant levels of water-band absorption were found around the primary targets WISE~0015$-$4615, WISE~1738$+$2732, and WISE~2019$-$1148. Table~\ref{t:candidates} summarises the relative astrometry and photometry of each candidate. All candidates were found at very wide angular separations ($\sim$25--35\arcsec), which would correspond to extremely large projected separations of 200--450~au if real.

Archival \textit{HST} data is available for two targets (WISE~0015$-$4615 and WISE~1738$+$2732), with images from programs GO~12972 (PI: Gelino), GO~12330 (PI: Kirkpatrick) and GO~16229 (PI: Fontanive) providing time baselines of at least 5 years to observations from our own program. Adopting astrometric solutions for the primaries from \citet{Kirkpatrick2021}, comparisons of measured astrometry at these additional epochs to that from our program demonstrated both candidates to be background sources, as shown in Figures~\ref{f:W0015} and \ref{f:W1738}.

The last candidate identified from its water absorption levels has no other observational epoch in the \textit{HST} archive and could not be astrometrically confirmed or refuted in the same manner.  
Based on the \textit{HST} work from \citet{Fontanive2018} (see Figure~\ref{f:SpT_Wabs}), the observed F127M$-$F139M maximum colours of the identified source suggest that the candidate is not compatible with a background main sequence star or early-type M-L brown dwarf --which do not show such deep water absorption levels-- and would instead have a $\geq$T2--T1 spectral type if a star-like object in the galaxy. 
Ultracool brown dwarfs like our primaries and any potential candidate companion to our targets are far too cold and red to be detected at visible wavelengths by Gaia, and even too faint at near-infrared wavelengths to be detected in 2MASS.
WISE~2019$-$1148 had been observed from the ground for spectro-photometric characterisation (Palomar/TSpec; \citealp{Mace2014}), but the field of view of the acquired observations do not cover the position of the extremely wide candidate. We did not find neither the primary nor the candidate in CFHT/MegaCam and SDSS archival images either. While detecting the source at optical wavelengths would have ruled out the candidate as real a ultracool companion, non-detections are inconclusive.
We were also unable to detect the identified source near WISE~2019$-$1148 in public WISE data, both assuming comoving and stationary background positions relative to the primary, and the source is not listed in the CatWISE and UnWISE catalogs \citep{Schlafly2019,Eisenhardt2020}. However, its absolute F127M magnitude of F127M = $22.47\pm0.17$, if at the distance of the primary, would place it at the faint end of currently-known Y1$-$Y2 dwarfs with existing F127M \textit{HST} data, like the much closer WISE~0350$-$5658 (6\,pc) and WISE~2354$+$0240 (8\,pc) \citep{Kirkpatrick2012,Schneider2015,Kirkpatrick2021}. Given the farther distance to WISE~2019$-$1148 (12.5\,pc), similar colours as these objects for the candidate around WISE~2019$-$1148 would translate to an apparent magnitude of $W2\sim16-17$~mag, potentially below the 16.5-mag threshold of the deep CatWISE program \citep{Eisenhardt2020}, which led to the discoveries of the faintest-known Y dwarfs to date \citep{Meisner2020a,Meisner2020b}. While non-detections of that source in these IR archives are hence inconclusive regarding the possible bonafide nature of the candidate, existing \textit{Spitzer} data of the primary (Program~70062, PI: Kirkpatrick) show no signs of infrared detections of the object either, despite a significantly deeper sensitivity in the $ch2$ channel (down to $ch2\sim$17--18~mag; \citealp{Martin2018,Kirkpatrick2019,Kirkpatrick2021}), where a Y-type dwarf of the expected brightness should be detected at 4.5~$\micron$. 

We therefore conclude that the source is unlikely to be a real companion and is instead more probably a background source, possibly an extra-galactic contaminant, which have previously been found to be sources of false positives to the water-band approach (e.g., \citealp{Fontanive2018}). Additional observations will be required to more confidently establish the true nature of the identified source. All selected candidates were hence refuted, and we conclude that no companion was detected in this survey around the 33 late-T and Y dwarfs observed by our \textit{HST} program.

\subsection{Survey Completeness}
\label{completeness}

In order to assess the completeness of our binary companion search, we made use of Artificial Star Tests similar to those described in Section~\ref{ASTs}, but with a focus around the primary brown dwarfs and more relevant regions of the spatial and magnitude parameter space.
For each target, we added $2\times10^6$ artificial stars across the full images, split into 4 sets. In all cases, we adopted flat distributions in the colour-magnitude plane, and used uniform radial distribution in position centred around the primary brown dwarf (i.e., giving the same number of sources at all radii, and uniformly distributed in position angle at a given radius).
A first set of $4\times10^5$ sources were added from a uniform radial distribution extending from 20 pixels of the brown dwarf target to the edge of the image, and with instrumental magnitudes drawn from uniform distributions between $-5$ and $+6$ in both F127M and F139M (Vega magnitudes of 18.55--29.55 in F127M and 18.25--29.25 F139M). 
A second set of $6\times10^5$ artificial stars was then added over the same radial distribution, beyond 20~pixels of the science target, but with instrumental magnitudes taken in the range $-2$ to $+2$ for each filter (Vega magnitudes of 21.55--25.55 in F127M and 21.25--25.25 in F139M), in order to further explore the magnitudes over which the completeness abruptly decreases. 
To complement these 1 million artificial stars away from the primary, we added another 1 million stars on uniform radial distributions between 1 and 20 pixels, so as to investigate in more details the region directly around the brown dwarf (inner $\sim$2.5\arcsec): a third test added $4\times10^5$ stars inside 20 pixels with instrumental F127M and F139M magnitudes uniformly drawn between $-5$ and $+6$, and a fourth test of $6\times10^5$ stars was drawn from instrumental magnitudes between $-2$ and $+2$. 

The source selection for the binary search in Section~\ref{comp_search} was based entirely on F127M detections, where we placed thresholds to select well-measured F127M sources independently of the various adopted categories of F139M detections (well-measured, poorly-measured but detected, or dropout), that were each considered and looked into separately. We therefore only considered the F127M ASTs to gauge the completeness of our analysis, although the corresponding F139M images were still used as well.
We applied the criteria and selection threshold defined in Section~\ref{ASTs} to assess which of the 2 million injected artificial stars were successfully recovered, and estimate our sensitivity and recoverability levels in each F127M image. In addition to these, our source selection procedure in the binary search (Section~\ref{comp_search}) included an additional requirement for sources to be at least 1.5 magnitudes brighter in F127M than the local background magnitude in F139M, in order to ensure a minimally meaningful magnitude drop between the two filters. We therefore enforced this extra constraint in the source selection of the ASTs as well, so as to comprehensively determine the completeness of the binary search that was performed here.

After running the source selection procedures on the 4 sets of ASTs, determining whether each artificial star was counted as well retrieved in F127M or not, we defined a grid in separation vs. magnitude space, allowing us to obtain the recoverability rate of the injected stars in each grid cell.
We used a separation bin size of 20 pixels at outer radii (beyond 20 pixels from the brown dwarf target) out to 1000-pixel separations, and refined the grid to 1-pixel resolutions for the inner 20 pixels. In terms of magnitudes, we used 0.1-mag bins from $-2$ to $+2$ mag where the completeness changes significantly, and bin sizes of 0.5~mag outside that range. The numbers of artificial stars injected in the 4 sets described above were specifically chosen to ensure having at least around 500 added stars per grid cell with these varying resolutions. Each star was placed in its corresponding separation-magnitude grid cell based on its input position and F127M instrumental magnitude, and the final fraction of stars well recovered by the algorithms was computed. 

\begin{figure}
    \centering
    \includegraphics[width=0.5\textwidth]{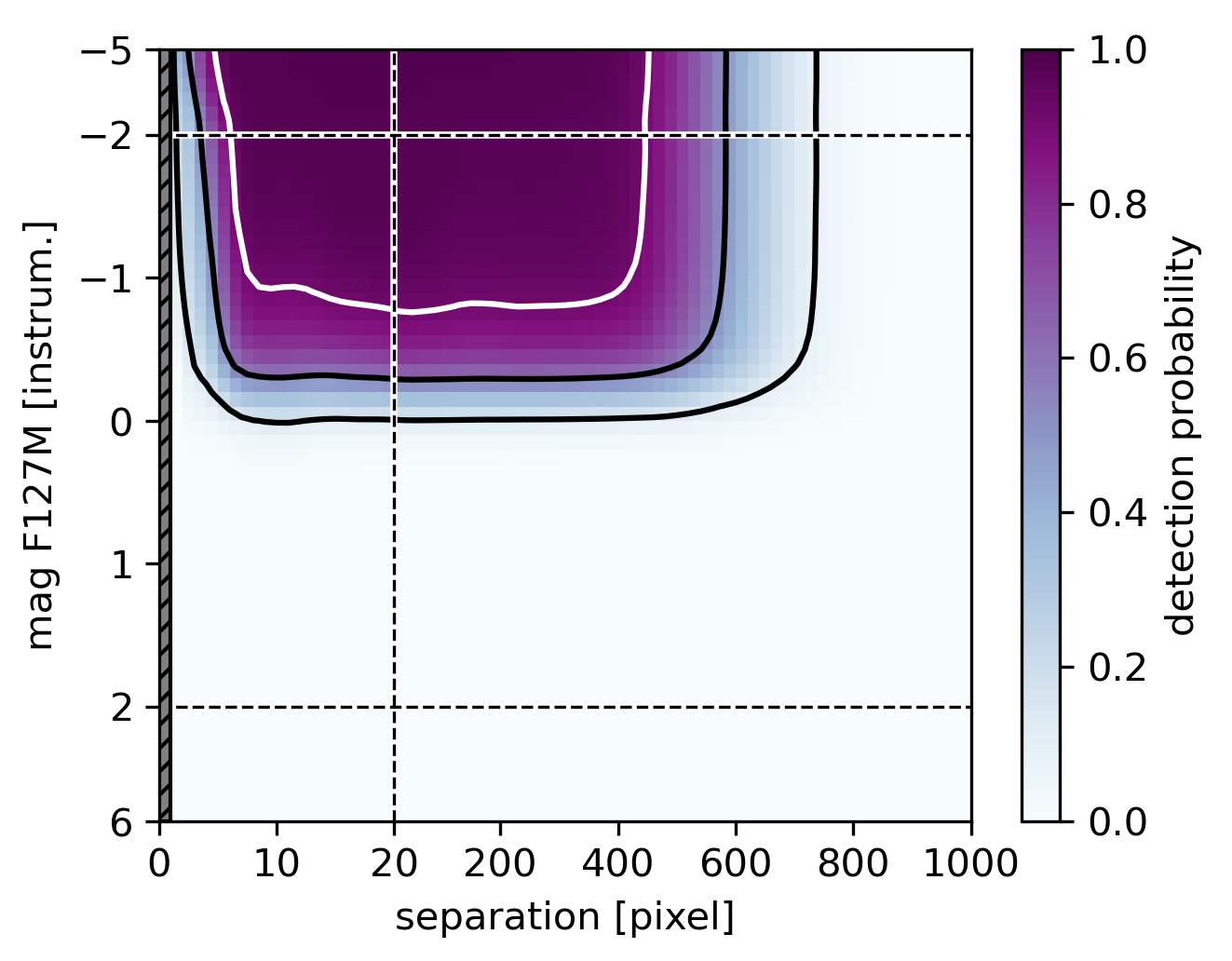}
    \caption{Average detection probability map for all 33 targets, based on the recoverability rate of artificial stars in each separation-magnitude bin. The white line indicates the 90\% contour, while the black lines show the 10\% and 50\% level of completeness. The grey shaded area on the left represents the 1-pixel inner  working angle. Axes are plotted in unit grid cells to highlight the key parts of the parameter space, and the linear scales change at the dotted lines, with a different resolution within and beyond 20~pixels, and inside and outside the $-2$ to $+2$ instrumental magnitude range. The fraction of a circle of given separation that lands in the image is taken into account in the final recoverability rate, resulting in a drop in the completeness at wide separations. }
    \label{f:completeness_grid}
\end{figure}

\begin{figure*}
    \centering
    \includegraphics[width=0.95\textwidth]{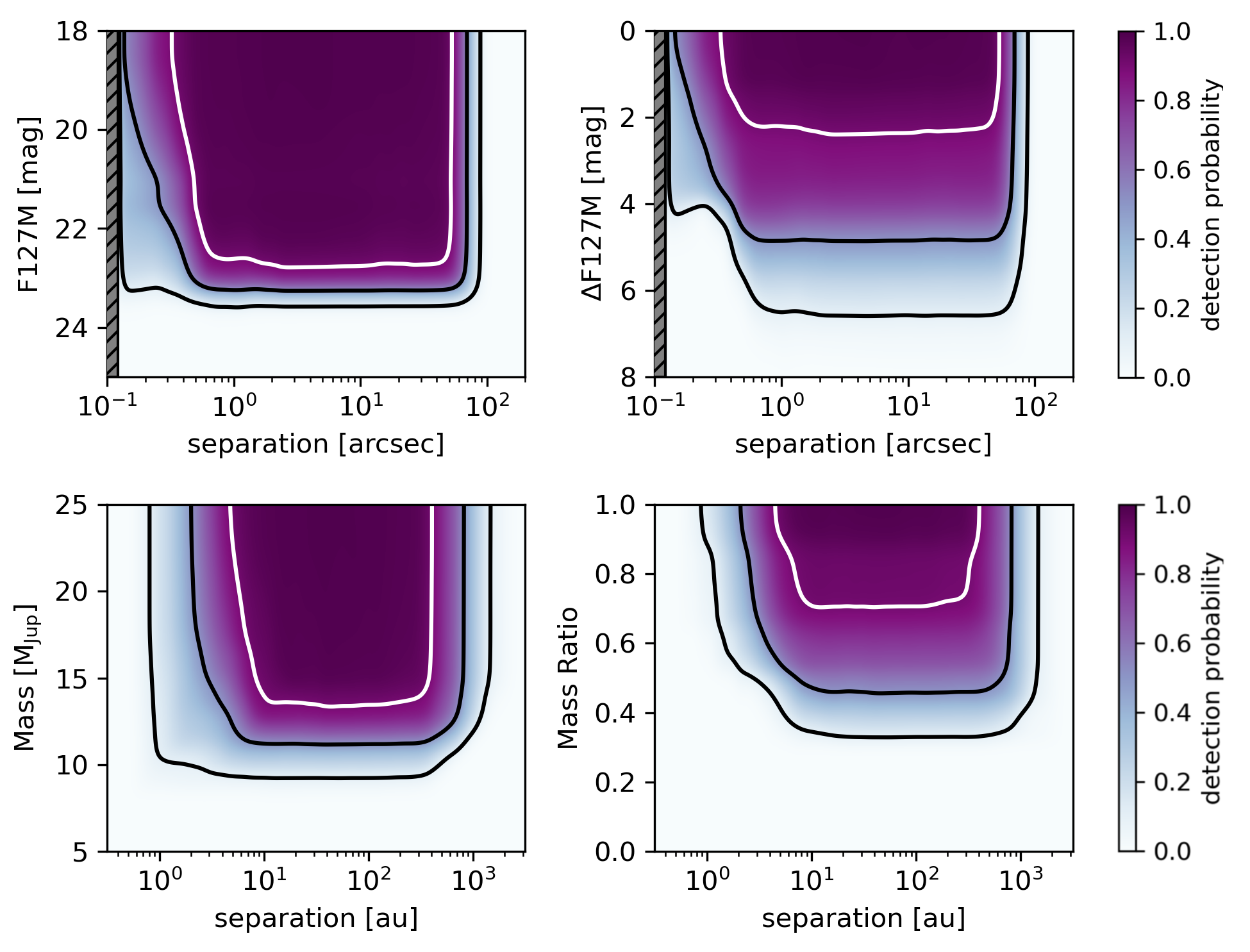}
    \caption{Detection probability maps for the full survey in observed (upper panels) and physical (lower panels) parameter space. The left panels show inherent companion quantities in the y-axis, while the variables plotted in the right panels are relative to the brown dwarf primaries. The black and white contours mark the 10\%, 50\% and 90\% completeness levels.}
    \label{f:all_completeness}
\end{figure*}

At separations larger than the closest edge of the image from the primary, a fraction of the considered annulus falls outside the field of view of the image, which means that the completeness of the program to companions on such separations should decrease accordingly. Nonetheless, the ASTs were made so that an equal number of sources would fall inside the images at any given radius from the target. This should be corrected in order to accurately reflect the completeness of the survey to the widest companions (e.g., an average recoverability rate of 0.8 at a separation where 50\% of the annulus falls outside the FoV, implies a true completeness of $0.8\times0.5 = 0.4$). Therefore, to extend the analyses beyond the radius that first reaches an edge of the image, we estimated for each separation the fraction of the annulus that fell within the FoV, and used that information to scale the obtained recoverability rates from the ASTs at that separation. This effect is reflected in the completeness grid in Figure~\ref{f:completeness_grid}, where the recoverability fraction starts to drop at outer separations (beyond $\sim$500~pixels for the averaged grid of the full sample). 

The added criterion on the minimum measurable magnitude drop between F127M and F139M for selected sources pushes the sensitivity down to instrumental magnitudes between 0 and $-1$, even though fainter sources not fulfilling these constraints are detectable in F127M  down to instrumental magnitudes of $+1$ to $+2$. Nonetheless, the extra selections applied in our binary search ensure that we focus our search on the regions of the parameter space with high-completeness levels, as demonstrated in Section~\ref{ASTs}, and most importantly, where very few poorly-measured sources are expected in the output of the algorithms.

Using the derived zeropoint value (ZP$_\mathrm{F127M}$ = 23.55) and the WFC3/IR pixel scale, the individual completeness grid for each brown dwarf in the sample was then converted from instrumental to Vega F127M magnitudes and $\Delta$F127M magnitude differences to the primaries, and transformed from pixels to arcseconds in separation. The distance to each target was subsequently used to convert from apparent to absolute F127M magnitude, and from projected angular separation in arcseconds to physical binary separation in astronomical units.
Finally, the ATMO2020 CEQ models \citep{Phillips2020} were used to derive corresponding masses at median ages of 5~Gyr, as was done in Section~\ref{BD_masses}, and we derived corresponding mass ratio grids from the primary masses estimated at the same age. 
Since conversions of the original grid axes are dependent on the individual distance and mass of each target, we recombined the final completeness maps by mapping them onto common, high-resolution master grids using 2-dimensional interpolations.

Figure~\ref{f:all_completeness} shows the final combined grids in the observed and physical parameter space. Beyond 0.5\arcsec ($\sim$4~pixels), we reach magnitude differences of $\Delta$F127M of 5~mag around 50\% of the sample. The wide spread of $>$4~mag in achieved contrasts across the sample (top right panel) is mostly due to the range in the primaries' brightness, since only $\sim$1~mag difference is seen in companion apparent magnitude (top left). These background-regime sensitivities correspond to companion mass limits of 11~M$_\mathrm{Jup}$ at the 50\% level, and 13.5~M$_\mathrm{Jup}$ at the 90\% from $>$5--10~au (bottom left), or mass ratios $q$ of 0.45 (50\%) and 0.7 (90\%) (bottom right). With targets distances ranging from $<$5~pc to 30~pc, the 1-pixel inner working angle (121~mas; shaded regions in the upper panels) corresponds to physical separations between 0.5--3.5~au across the sample.
As a results, in the contrast-limited regime within $\sim$0.5\arcsec, we are sensitive to equal-mass companions from 1~au for the best 3 targets, from 2~au around half of the targets, and starting at 5~au for over 90\% of cases. 

While the model-based mass conversion for the lower left panel depends on the adopted age for the systems, the resulting mass ratios in the lower right panel show little age dependence. Indeed, when considering different ages, the mass sensitivity limits were found to shift from 11~M$_\mathrm{Jup}$ to 7~M$_\mathrm{Jup}$ and 14~M$_\mathrm{Jup}$ for the 50\% level at 2~Gyr and 8~Gyr, respectively, and from 13.5~M$_\mathrm{Jup}$ to 9~M$_\mathrm{Jup}$ and 17~M$_\mathrm{Jup}$ for the 90\% contours for the same ages. On the other hand, working in mass ratio space, with primary masses computed at the same ages, results in changes of $\sim$1\% in the corresponding $q$ levels. This makes mass ratio an advantageous variable for an analyses of binary properties, providing an almost age-independent space to work in, as had been showed in \citet{Fontanive2018}. We therefore focus on that parameter space for the statistical analysis in Section~\ref{stats}. The detection probability map, in physical projected separation against mass ratio, contains 1000 values uniformly distributed in log space between $10^{-0.5}$~au and $10^{3.5}$~au for the separation axis (0.31--3100~au), and 500 values in mass ratio linearly spaced between 0 and 1.

\subsection{Constraints on Binary Frequency}
\label{stats}

\begin{figure}
    \centering
    \includegraphics[width=0.45\textwidth]{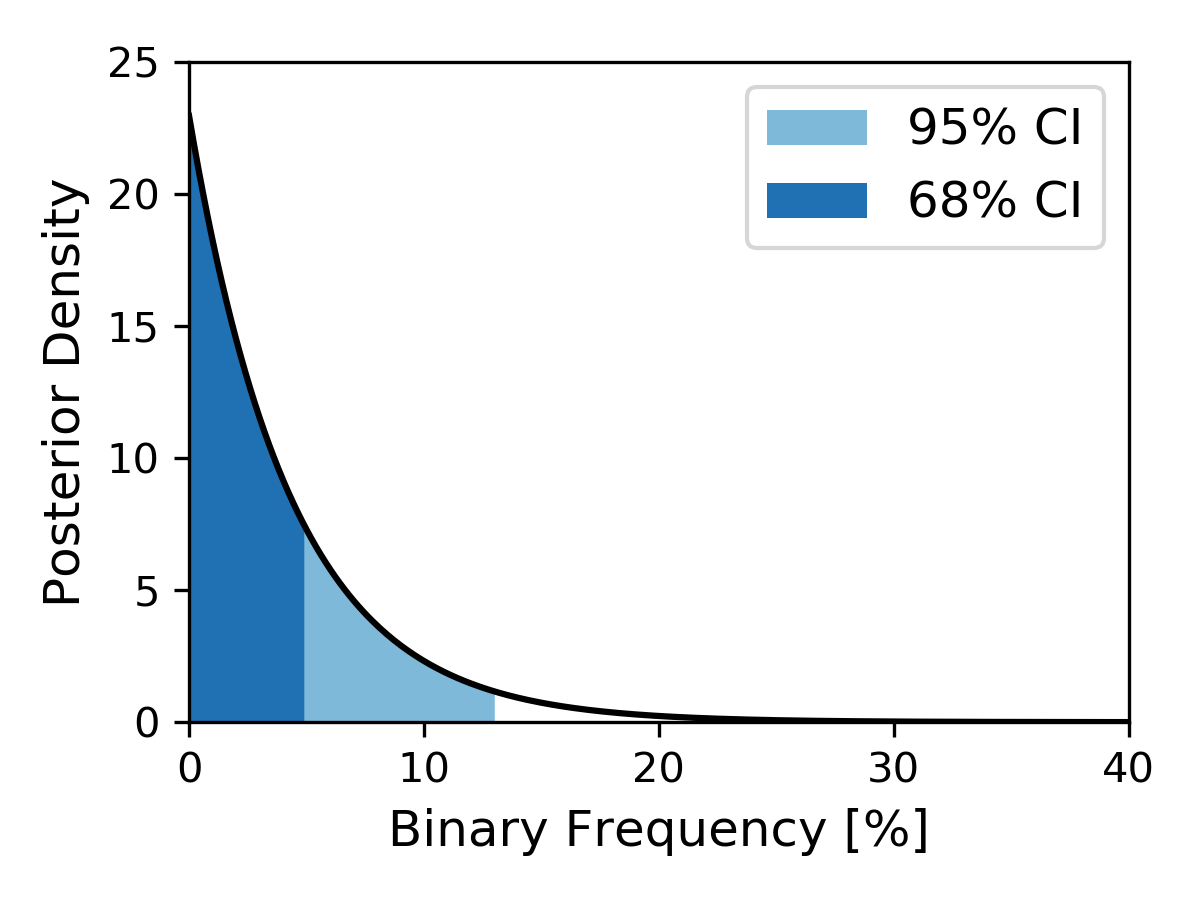}
    \caption{Posterior probability distribution of the binary frequency $f$ on separations 1--1000~au and mass ratios 0.4--1, for our observed sample of 33 targets. The shaded areas represent the 1-$\sigma$ (deep blue) and 2-$\sigma$ (light blue) intervals of confidence, respectively.}
    \label{f:frequency}
\end{figure}

With a statistically significant sample of 33 ultracool brown dwarfs and a deep sensitivity to low-mass objects, the lack of detected true companion in this survey allows us to place stringent upper limits on the occurrence of binary companions around late-T and Y dwarfs. To do this, we adopt the Bayesian statistical framework from \citet{Fontanive2018}, a formalism initially developed to constrain the binary properties of late-T dwarfs, in the predecessor survey to the one conducted here. Without any detections, we are unable to place new observational constraints on the binary orbital separation or mass ratio distributions of this population. We therefore have to rely on existing results for the shape of the underlying binary population. 

In exoplanet surveys around stars, the output of planet formation theories, in the form of population synthesis models \citep{Forgan2018,Emsenhuber2021}, are typically available for direct comparisons of observed populations to theoretical predictions (e.g., \citealp{Vigan2021}). 
Unfortunately, such simulations are very limited for brown dwarf populations, and rarely extend down to the very low-mass end of the substellar regime \citep{Bate2014}, populated by isolated T and Y dwarfs like the ones studied here. 

In the absence of such theoretical expectations for the distributions of orbital periods and binary mass ratios, we adopt the results from \citet{Fontanive2018}. That study derived a binary frequency of $f = 5.5^{+5.2}_{-3.3}$\% for T5–Y0 brown dwarfs over separations of 1.5--1000~au, for an overall binary fraction of $8\pm6$\%.
Modelling the projected separation as a lognormal distribution and the mass ratio as a power law, they found a peak in separation at $\rho = 2.9^{+0.8}_{-1.4}$~au with a logarithmic width of $\sigma = 0.21^{+0.14}_{-0.08}$, and mass ratio distribution heavily skewed towards equal-mass systems, with a power-law index of $\gamma = 6.1^{+4.0}_{-2.7}$. We followed the procedures and used the Markov chain Monte Carlo (MCMC) tools from \citet{Fontanive2018,Fontanive2019}, based on the \texttt{emcee} python implementation of the affine-invariant ensemble sampler for MCMC \citep{Foreman-Mackey2013}.
We generated binary populations from the separation and mass ratio distributions described above, and combined them with our survey detection probability map generated in Section~\ref{completeness} to derive an upper limit on the binary rate $f$ of our observed sample given our null detection. We focused on the region of the parameter space between $\rho = 1-1000$~au and $q = 0.4-1$, down to the limits of our binary search sensitivity level (see Figure~\ref{f:all_completeness}). Adopting a uniform prior between 0 and 1 for $f$, we found a binary frequency of $f < 4.9$\% ($< 13.0$\%) at the 1-$\sigma$ (2-$\sigma$) level. Figure~\ref{f:frequency} shows the obtained posterior distribution for the binary fraction with the corresponding 68\% and 95\% confidence intervals indicated by the shaded regions.

\section{Discussion}
\label{discussion}

\begin{figure}
    \centering
    \includegraphics[width=0.45\textwidth]{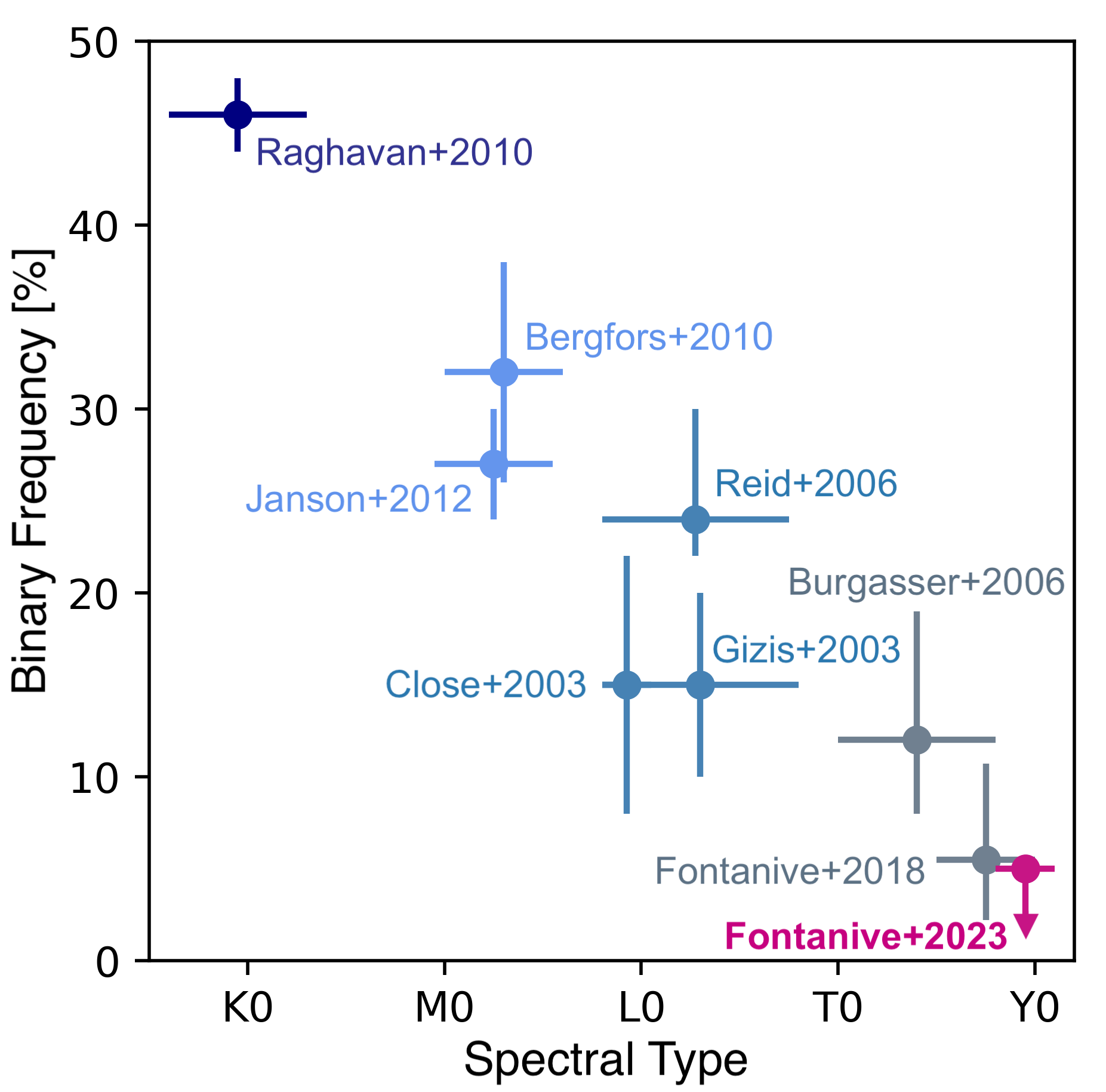}
    \caption{Binary fraction as a function of spectral type for stars and brown dwarfs in the field, showing the continuous decrease in multiplicity rate with later type. Our measured upper limit for T8--Y1 dwarfs is plotted in magenta. All measurements come from surveys probing similar binary separations to our programme,  $>1$--3~au, except for the \citet{Raghavan2010} result for G and K stars. Data from \citet{Raghavan2010,Bergfors2010,Janson2012,Close2003,Gizis2003,Reid2006,Burgasser2006,Fontanive2018}.}
    \label{f:BF_vs_SpT}
\end{figure}

Figure~\ref{f:BF_vs_SpT} compares our obtained constraint in binary fraction to values measured for earlier-type, more massive field brown dwarfs and stars. Apart from the G-K dwarf binary rate \citep{Raghavan2010}, all frequencies were derived from surveys probing systems with separations $>$1--3~au, comparable to our programme, making the comparison in multiplicity fractions directly relevant. Our measured upper limit of $f < 4.9$\% (68\% confidence interval) for the frequency of systems on separations of 1--1000~au and mass ratios between 0.4--1 is in good agreement with previous results derived in \citet{Fontanive2018} for a sample of slightly earlier-type T5-Y0 brown dwarfs ($f = 5.5^{+5.2}_{-3.3}$\% beyond 1.5~au, for an overall binary fraction of $8\pm6$\%). These findings marginally confirm the idea that the decrease in binary frequencies with later type observed across the stellar and substellar regimes for the field population might continue throughout the substellar mass range down to the very lowest masses, as illustrated in Figure~\ref{f:BF_vs_SpT}. The continuity in this trend across the star-brown dwarf boundary would argue for a common origin between stars and brown dwarfs, although larger sample sizes and more stringent constraints are still needed to validate this result.

As brown dwarfs are incredibly faint, the methods most commonly used to search for binary and planetary companions to stars, like the radial velocity and transit methods, are typically unfeasible in the brown dwarf regime, in particular for ultracool late-type objects at the low-mass end like the targets studied here. Astrometry might provide a more viable alternative approach to search for companions to faint brown dwarfs (e.g., \textit{HST} Program GO~17080\footnote{\url{https://ui.adsabs.harvard.edu/abs/2022hst..prop17080B/abstract}}, PI Bedin), although very little work has been carried out on this side and no systems have been reported this way so far. Direct imaging thus remains the primary option available to search for companions to brown dwarfs, but is limited to relatively high-mass ratios and is only sensitive to the outer parts of the separation parameter space. Brown dwarf binaries appear to be in tighter orbital configurations than their more massive stellar counterparts \citep{Burgasser2003,Burgasser2006}, consistent with the trend observed in binary separation distribution of multiple systems within the stellar regime. As this tendency extends across the brown dwarf mass range all the way down to the very lowest masses \citep{Fontanive2018}, the peak of the orbital separation distribution for the latest-type primaries approaches the resolving limit of most direct imaging instruments capable of observing brown dwarfs. 

For instance, in the few searches for companions to late-T and Y primaries conducted to date, most existing discoveries emerged from the surveys with the best inner working angles (Keck/NIRC2, \citealp{Liu2011,Dupuy2015}), which would have been (or were) missed by studies only able to access wider angular separations (e.g., \textit{HST}/WFC3, \citealp{Aberasturi2014,Fontanive2018}). As we are mostly sensitive to companions beyond few au, the lack of new detection in our \textit{HST} program is hence consistent with previous results of substellar binarity. Indeed, our survey's inner working angle coincides roughly with the peak around 3~au in orbital separation from \citet{Fontanive2018} for such systems, meaning that we are unable to detect at least half of potentially existing companions.
With the largest sample of Y dwarfs ever probed for multiplicity, our strong constraints of $f < 4.9$\% (1-$\sigma$) between 1--1000~au thus confirms that the binary rate of brown dwarfs continues to decrease with later spectral types all the way down to Y dwarfs on these separations. These results could also be consistent with a further shift of the binary separation peak to even smaller separations around Y dwarfs. The first and only binary system with a Y-type primary (WISE J033605.5$-$014350.4, which is not in our sample) was recently discovered with \textit{JWST}/NIRCam \citep{Calissendorff2023}, but required \textit{JWST}'s plate-scale (63~mas and 31~mas for the F480M and F150W filters used, respectively) to detect the $\sim$84~mas (0.97~au) Y$+$Y binary. The rare current discoveries therefore strongly indicate that spatial resolutions reaching sub-au separations are likely needed to uncover more members of this binary population. Future surveys exploring the inner regions of the separation space will be crucial to provide a definitive test of whether the observed peak around 2--3~au for T and Y dwarfs \citep{Fontanive2018} is real, or if it is an artefact of incompleteness on shorter separations \citep{BardalezGagliuffi2015}, with a true peak lying at even tighter separations. The 3--5~$\mu$m sensitivity and high angular resolution of \textit{JWST} will allow us to study these objects in larger samples at closer separations (e.g. \textit{JWST} Cycle 1 program GO~2473\footnote{\url{https://ui.adsabs.harvard.edu/abs/2021jwst.prop.2473A/abstract}}, PIs Albert \& Meyer).

With an excellent sensitivity and completeness to companions on wide orbital separations (outside 0.2--0.5\arcsec, or 1--10~au, depending on the targets' distances and brightness), our survey robustly confirms that wide companions are extremely rare in the Galatic field around the lowest-mass systems. The binaries discovered by \citet{Liu2012} with separations of 8--15~au thus remain highly anomalous and the only examples of binaries on outer Solar System scales with such low-mass primaries. We also know that yet wider systems with separations of tens to hundreds of au do form around similar mass primaries, since they are observed in young star-forming regions and moving groups \citep{Chauvin2005,Todorov2010,Fontanive2020}. The fact that no analogue to these extremely wide, low-mass binaries has ever been uncovered at evolved ages, despite a high sensitivity to such companions around nearby old objects, indicates that such systems do not survive to field ages \citep{Burgasser2006,Biller2011}. Our results, with no detection of wide companions out of 33 observed objects, reinforce the idea that the widely-separated binaries with very low-mass primaries identified in young associations have no counterparts among isolated objects in the Solar neighbourhood.

\section{Conclusions}
\label{conclusions}

We conducted an imaging search for low-mass binary and planetary companions around 33 nearby brown dwarfs with spectral types of T8--Y1, using WFC3/IR observations from the Hubble Space Telescope. With 9 Y-type primaries, our observed sample is the largest subset of such very late-type, ultracool brown dwarfs investigated for multiplicity to date. Our deep \textit{HST} observations provide the first detections of late-T dwarfs in the water-absorption band at 1.4~$\micron$, providing new observational constraints on the depth of this spectroscopic feature at the coldest temperatures. We derive new estimates of our targets' luminosities, temperatures and masses from evolutionary models, using our \textit{HST} near-infrared photometry combined with \textit{Spitzer} data in the infrared.

We found no evidence for wide binary companions in our survey, despite an excellent sensitivity and completeness from a few au and down to secondary masses of 10--15~M$_\mathrm{Jup}$. Three candidates were identified in our procedures for companion selection based on water-band absorption levels, but two were ruled out from their stationary background positions relative to the primaries in archival \textit{HST} data, while the third source was rejected a bonafide companion based on its non-detection in deep infrared \textit{Spitzer} images, where such a cold companion should be bright.
We were nonetheless able to place stringent constraints on the binary rate $f$ of these objects from our null detection, and measured an upper limit of $f < 4.9\%$ at the 1-$\sigma$ level ($< 13.0$\% at the 2-$\sigma$ level) over the separation range 1--1000~au and for mass ratios between 0.4--1. This is consistent with previous results, and reaffirms that the decrease in multiplicity fraction with later spectral types seen among stars and more massive brown dwarfs might persist all the way down to the latest-type objects.

Our survey confirms that wide companions are extremely rare around the lowest-mass and coldest isolated brown dwarfs, even though counterpart systems are observed at young ages, indicating that such systems do not survive as bound components to field ages. If our targets have binary companions, they would likely be within our inner working angle of $\sim$1--5~au, on separations around or inside the currently-observed peak in binary separation for T and Y dwarfs. 
Future programs that explore the inner regions of the separation space will be crucial to identify more members of this binary population and provide a definitive test of the observed separation peak.

\section*{Acknowledgements}

CF acknowledges support from the Trottier Family Foundation and the Trottier Institute for Research on Exoplanets through her Trottier Postdoctoral Fellowship. This work was funded by the Trottier Institute for Research on Exoplanets and the Center for Space and Habitability, and partly carried out within the framework of the National Centres of Competence in Research PlanetS supported by the Swiss National Science Foundation.
LRB acknowledges partial support by MIUR under PRIN programme \#2017Z2HSMF and by PRIN-INAF 2019. 
This survey is based on observations with the NASA/ESA Hubble Space Telescope, obtained at the Space Telescope Science Institute, which is operated by AURA, Inc., under NASA contract NAS 5-26555. These observations are associated with programme 15201. Some of the data presented in this paper were obtained from the Mikulski Archive for Space Telescopes (MAST).

\section*{Data Availability}

All observational data used in this work are publicly available. The reduced stacked images from this work will be supplied upon request to the corresponding author.



\bibliographystyle{mnras}
\input{main.bbl}





\bsp	
\label{lastpage}
\end{document}